\documentclass[11pt,letterpaper]{article}
\pdfoutput=1
\usepackage{jheppub}
\usepackage{color}
\usepackage{graphicx}

\usepackage{verbatim}
\usepackage{amsmath}
\usepackage{amssymb}
\usepackage{url}
\usepackage{xspace}
\usepackage{tabularx}
\usepackage[makeroom]{cancel}

\usepackage{multirow}

\usepackage{color}

\newcommand{\beq}{\begin{equation}}
\newcommand{\eeq}[1]{\label{#1}\end{equation}}
\def\beqa{\begin{eqnarray}}
\def\eeqa#1{\label{#1}\end{eqnarray}}
\newcommand{\eeqn}{\end{equation}}
\newcommand{\CR}{\notag \\}
\newcommand{\leqn}[1]{(\ref{#1})}

\def\bea{\begin{eqnarray}}
\def\eea{\end{eqnarray}}

\def\zero#1{\tilde{#1}}

\newcommand{\bspace}{\!\!\!\!}
\def\met{\mbox{$E{\bspace}/_{T}$}}
\def\N{{\cal N}}
\def\GeV{~{\rm GeV}}

\def\stacksymbols #1#2#3#4{\def\theguybelow{#2}
    \def\vp{\lower#3pt}
    \def\sp{\baselineskip0pt\lineskip#4pt}
    \mathrel{\mathpalette\intermediary#1}}

\def\intermediary#1#2{\vp\vbox{\sp
     \everycr={}\tabskip0pt
     \halign{$\mathsurround0pt#1\hfil##\hfil$\crcr#2\crcr
              \theguybelow\crcr}}}

\def\lsim{\stacksymbols{<}{\sim}{2.5}{.2}}
\begin{document}

\title{Clockwork Neutrinos}

\author{Sungwoo Hong,}
\author{Gowri Kurup,}
\author{and Maxim Perelstein}

\affiliation{Laboratory for Elementary Particle Physics, Cornell University, Ithaca, NY 14853, USA}

\emailAdd{sh768@cornell.edu}
\emailAdd{gk377@cornell.edu}
\emailAdd{mp325@cornell.edu}

\date{\today}

\abstract{
Clockwork (CW) mechanism can explain the smallness of neutrino masses without introducing unnaturally small input parameters.   
In this paper we study the simplest CW neutrino model, the ``uniform" clockwork, as well as a broader class of ``generalized" clockwork models.
We derive constraints on such models from lepton-flavor violating processes, as well as precision electroweak fits. These constraints allow excited CW neutrino states with masses of order 100 GeV -- 1 TeV, within reach of the LHC and proposed lepton colliders, as long as the input neutrino Yukawa coupling is of order $10^{-1}-10^{-2}$. We study collider phenomenology of these models. At the LHC, models with light ($\sim 100$~GeV) CW neutrinos can be discovered using the $3\ell+\met$ signature. Lepton colliders will be able to discover the CW neutrinos as long as they are within their kinematic range.
}


\maketitle

\section{Introduction}

Neutrino masses are at least six orders of magnitude smaller than the mass of the electron, and at least twelve orders of magnitude below the scale where all fermion masses are thought to originate, the electroweak scale. The most popular explanation for the smallness of neutrino masses is the see-saw mechanism. While simple and theoretically attractive, this mechanism depends crucially on violation of lepton number symmetry. At this time, there is no experimental evidence that lepton number is violated, and it is a logical possibility that this symmetry is exact (or broken only by gravitational interactions). In this case, neutrino masses must be Dirac, and an alternative to see-saw is required to generate hierarchically small neutrino masses. 

It is possible to generate small Dirac neutrino masses in models with extra dimensions of space. If the left-handed neutrino fields are localized to a brane, along with other fields charged under Standard Model (SM) gauge symmetries, while the right-handed neutrino field propagates in the bulk, the Dirac mass is suppressed by the geometric factor reflecting the small overlap between the left-handed and righ-handed wavefunctions. This idea has been realized in the context of large extra dimensions~\cite{ArkaniHamed:1998vp,Dienes:1998sb}, and in Randall-Sundrum setup~\cite{Grossman:1999ra}. 

Recently, a new mechanism for generating exponentially small couplings and masses has been proposed, the Clockwork (CW) Mechanism~\cite{Giudice:2016yja}. Among other applications of the Clockwork, it has been suggested that it can be used to generate the observed neutrino masses without hierarchically small parameters in the Lagrangian of the theory. The right-handed neutrino emerges from a chain of four-dimensional fields with nearest-neighbor interactions in the theory space, while the left-handed neutrino is localized in the theory space. The right-handed neutrino zero-mode is exponentially suppressed at the site where the left-handed field resides, leading to an exponentially suppressed mass. The idea is similar to that of extra-dimensional models, and some CW models may related to 5D constructions (with an appropriately chosen metric profile) by dimensional deconstruction~\cite{ArkaniHamed:2001ca}. In this paper, we will use the four-dimensional point of view.  

The goals of this paper are to further develop the idea of CW mechanism for small neutrino masses, and to explore its phenomenological consequences. (For previous work on CW mechanism applied to neutrino masses, see~\cite{Park:2017yrn,Ibarra:2017tju}.) In Section~\ref{sec:base}, we present the simplest model that realizes the CW mechanism for neutrinos, which we call the {\it uniform clockwork}. This follows closely the model originally proposed in Ref.~\cite{Giudice:2016yja}, but generalizes it to fully incorporate the three neutrino flavors of the SM, including flavor-mixing effects required by the observed neutrino oscillations. 
We also present analytic expressions for the spectrum and couplings of the ``excited" CW neutrino states. These expressions are obtained within a perturbation theory in the parameter $p$, proportional to the Yukawa interaction which couples the SM to the CW sector. Analytic perturbative expressions provide intuitive understanding of various phenomenologically important quantities. In Section~\ref{sec:constraints}, we discuss experimental constraints on this model from flavor-changing neutral current process $\mu\to e\gamma$ and precision electroweak fits, and delineate the parameter space allowed by the existing data. We find that the excited CW neutrino states may have masses around the weak scale, in the 100 GeV -- 1 TeV range, without violating any constraints, with Yukawa couplings of order $10^{-2}-10^{-1}$. In Section~\ref{sec:extended}, we discuss the ``generalized" CW models, a generalization of the uniform model which allows for site-dependent Dirac masses in the clockwork sector. (An interesting example, the case of randomly drawn masses, has been previously considered in~\cite{Craig:2017ppp}.) We give the general condition under which the generalized model produces a hierarchically small neutrino mass, and consider two explicit examples, ``Linear Clockwork" models. Spectra and couplings of the excited CW neutrinos in these models are qualitatively different from the uniform CW case. As in the uniform case, we consider experimental constraints on the linear CW models, and find that in one of the linear CW models (LCW1) weak-scale excited neutrinos are allowed. Finally, in Section~\ref{sec:coll}, we discuss collider phenomenology of the uniform CW and LCW1 models. We find that excited neutrinos can be produced at the LHC and the proposed next-generation electron-positron colliders with significant rates. We identify signatures of CW neutrino production at hadron and lepton colliders, and perform Monte Carlo studies of these signatures and their SM backgrounds. We find that with the current data set, the LHC does not yet have the sensitivity to this model. Some spectra with relatively light CW neutrinos can be probed at HL-LHC, but sensitivity decreases rapidly as the CW mass scale is increased. At lepton colliders, the situation is more promising, and the CW neutrinos can typically be discovered with realistic integrated luminosity as long as they are within the kinematic reach of a given collider. In Section~\ref{sec:conc}, we summarize our findings and discuss possible directions for future studies of CW neutrino phenomenology.                    

\section{Uniform Clockwork Neutrino Model}
\label{sec:base}

\begin{figure}[t!]
	\center
	\includegraphics[width=14cm]{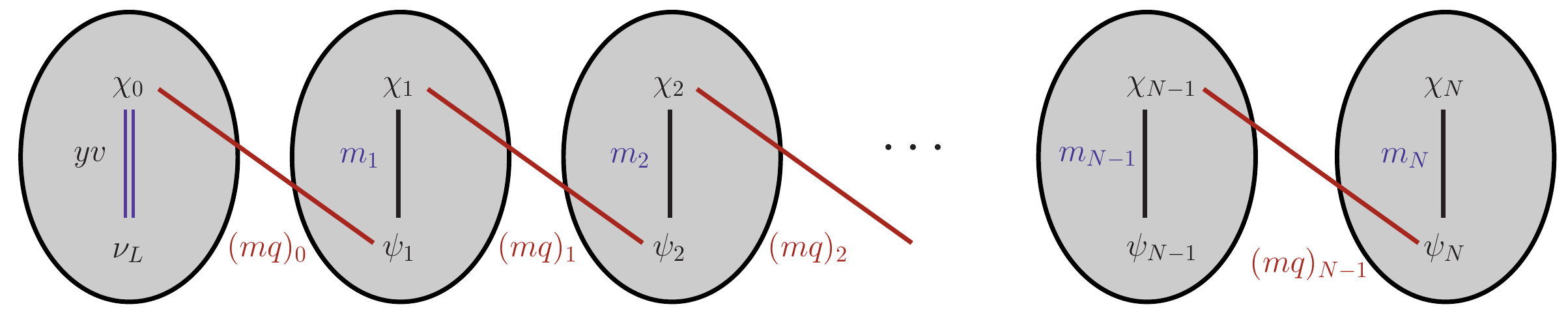}
	\caption{Pictorial representation of the clockwork sector with right-handed zero mode. Single solid lines denote Dirac masses, while the double solid line denotes the Yukawa couplings involving the SM Higgs boson $H$. In the uniform CW model, $m_i\equiv m$ and $q_i\equiv q$ for all $i$.}
	\label{fig:clockwork_fermion}
\end{figure}

The uniform clockwork neutrino model supplements the SM particle content with the following ``clockwork fields":

\begin{itemize}

\item  $N$ left-handed Weyl fermion fields $\psi_{i}$, where $i=1\ldots N$; 

\item $N+1$ right-handed Weyl fermions $\chi_{j}$, where $j=0\ldots N$.    

\end{itemize}

All clockwork fields are singlets under the SM gauge groups. The Lagrangian is 
\beq
{\cal L}_{\rm cw} = {\cal L}_{\rm kin} - m \sum_{i=1}^{N} \left( \psi_{i}^\dagger \chi_{i} - q \, \psi_{i}^\dagger \chi_{\tiny{(i-1)}} + {\rm h.c.}\right)\,,
\eeq{Lcw}  
where $m$ is the mass parameter (``clockwork mass") and $q$ is a dimensionless number of order one. We will assume $q>1$, which, as we will see below, results in exponential suppression of neutrino mass. In this paper, we will primarily consider $m$ at the weak/TeV scale, motivated by the desire to accommodate the observed neutrino masses without introducing new scale hierarchies. In the uniform model, $m$ and $q$ are the same for each term in the sum; the model has a ``translational" symmetry, $i \to i+1$, broken only by the edge terms. The particle content and mass terms of the model are represented pictorially in Fig.~\ref{fig:clockwork_fermion}, where we represent each pair of left- and right-handed fields as a ``site" (a gray circle), and each non-diagonal mass term as a ``link" (a red line). 

To incorporate three generations of neutrinos, we promote each of the clockwork fields to a flavor triplet, $\psi_{i\alpha}$ and $\chi_{j\alpha}$, $\alpha=1\ldots 3$, and assume that all mass terms in Eq.~\leqn{Lcw} are diagonal in flavor space. With this assumption, the model has a global flavor $SU(3)_{\rm cw}$ symmetry under which all clockwork fields transform in fundamental representation. 

The clockwork sector is coupled to the SM through the Yukawa coupling connecting the ``extra" right-handed clockwork fermion $\chi_0$ to the SM lepton doublet $L$: 
\beq
{\cal L}_{\rm Yuk} = Y^{\alpha\beta} \chi_{0\alpha}^\dagger (H \cdot L_\beta) + {\rm h.c.}
\eeq{LYuk}
where $H$ is the SM Higgs doublet, and $Y$ is the matrix of Yukawa couplings. The Yukawa coupling explicitly breaks the $SU(3)_{\rm cw}\times SU(3)_L$ flavor symmetry; by construction, it is the only source of such breaking. In this sense, the model incorporates minimal flavor violation in the neutrino sector. This choice is motivated by non-observation of lepton flavor violating (LFV) processes, and is advantageous from the point of view of minimizing experimental constraints. We will work in a basis where the Yukawa matrix is diagonal: 
\beq
Y={\rm diag}\,(y_1, y_2, y_3).
\eeq{Ydiag}
This assumption entails no loss of generality, since one can always perform flavor $SU(3)$ rotations $\psi_i \to V_{\rm cw} \psi_i$, $\chi_j \to V_{\rm cw} \chi_j$, and $L\to V_L L$, to diagonalize $Y$ without affecting the Lagrangian in Eq.~\leqn{Lcw}. Note that in this basis, lepton couplings to the SM $W$ boson are not flavor-diagonal; their flavor structure is described by the usual PMNS matrix.      

\begin{figure}[t!]
	\center
	\includegraphics[width=7.5cm]{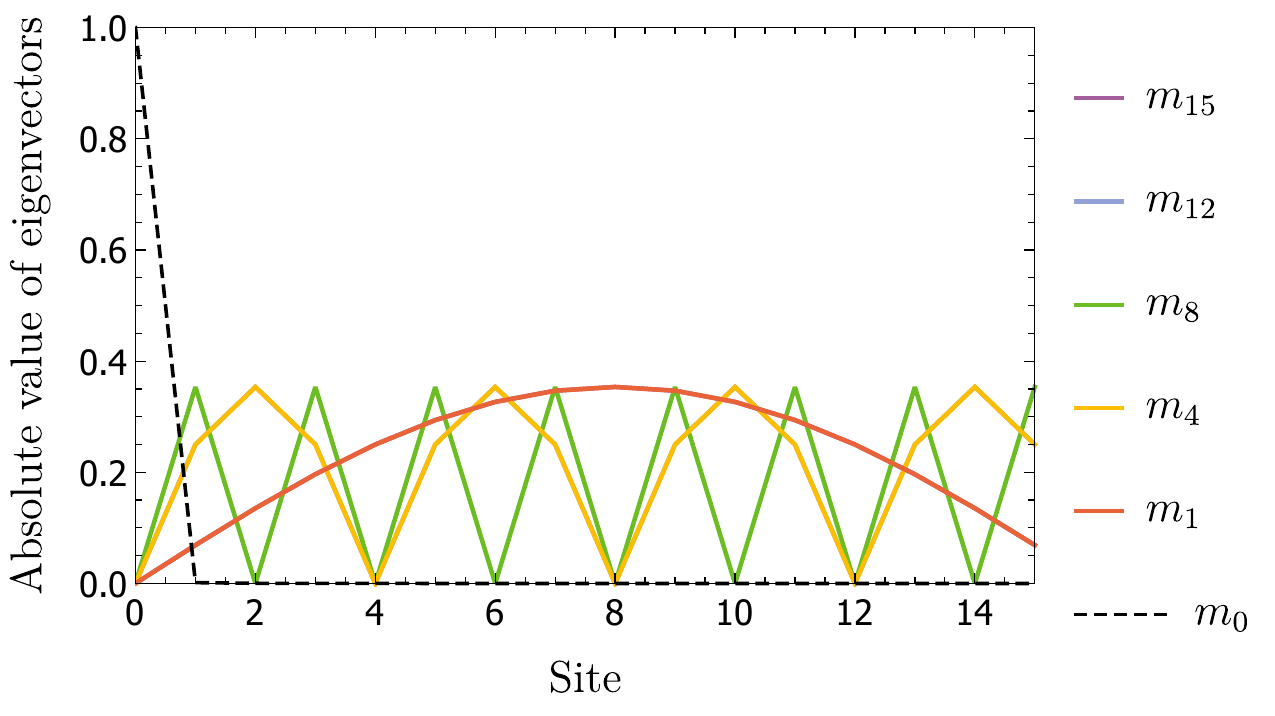}
	\includegraphics[width=7.5cm]{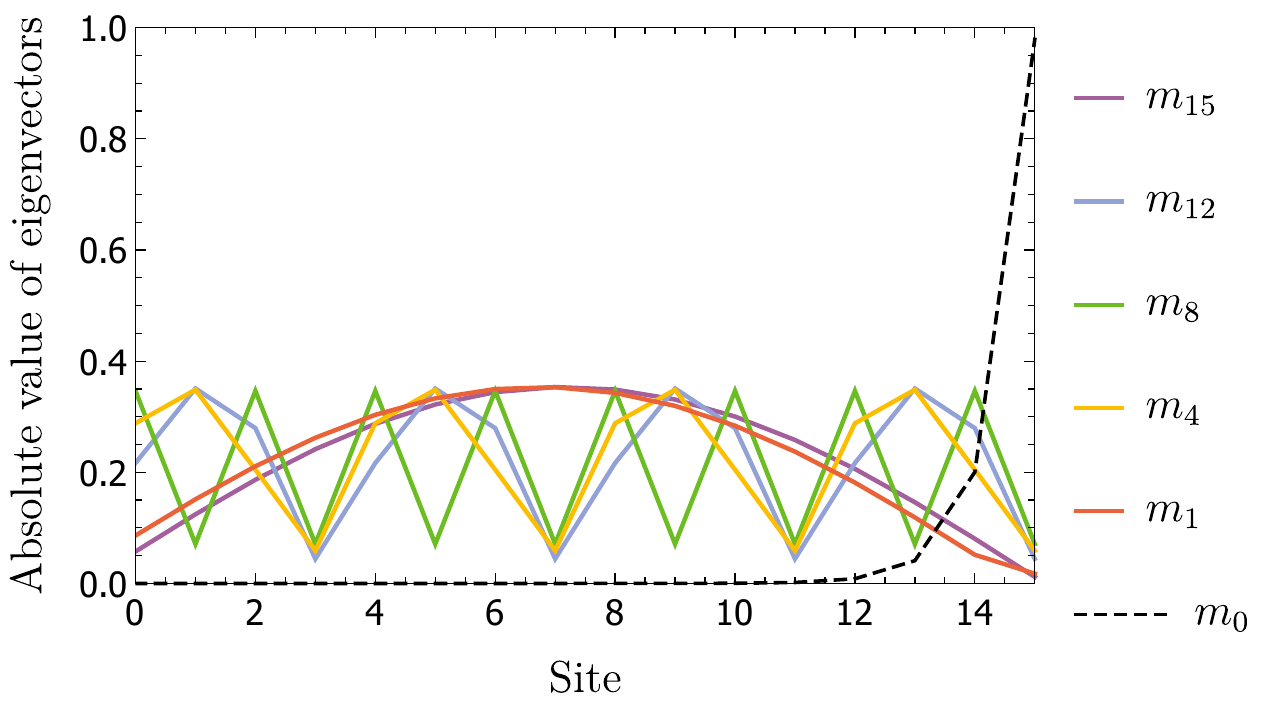}
	\caption{Composition of the left-handed (left panel) and right-handed (right panel) mass eigenmodes in terms of the original clockwork fields in the uniform clockwork model with $N=15$, $m=v$, $q=4.887$, and $y=0.01$.}
	\label{fig:localization}
\end{figure}

To understand the mass spectrum of the model, first consider the limit $Y=0$. Defining ``neutrino vectors"
\beqa
\Psi &=& (\nu_L, \psi_1, \psi_2, \ldots \psi_N)^T; \CR
X &=& (\chi_0, \chi_1, \chi_2, \ldots, \chi_N)^T,
\eeqa{bigfields}  
the mass term has the form $\Psi^\dagger \zero{M} X$ + h.c., where the mass matrix is given by
\beq
\zero{M} = m
\begin{pmatrix}
	0 & 0 & 0 & \cdots & 0 & 0 \\
	-q & 1 & 0 & \cdots & 0 & 0 \\
	0 & -q & 1 & \cdots & 0 & 0 \\
	\vdots & \vdots & \vdots & \ddots & \vdots & \vdots \\
	0 & 0 & 0 & \cdots & -q & 1
\end{pmatrix}.
\eeq{eq:fermion_general_mass_matrix}
(Here and below, tildes indicate the $Y=0$ limit.) The mass matrix can be diagonalized by a pair of unitary rotations, $\zero{U}_L$ and $\zero{U}_R$, with the mass eigenstates $\zero{\N}_L$ and $\zero{\N}_R$ given by 
\beq
\Psi = \zero{U}_L \zero{\N}_L,~~~~~X = \zero{U}_R \zero{\N}_R. 
\eeq{Mass_eigenstates}
Translational symmetry of the model allows for exact, analytic diagonalization of the mass matrix. Since $\det \zero{M}=0$, there is a massless eigenstate, the zero-mode. The spectrum of massive modes is given by 
\beq
\zero{m}_k = \lambda^{1/2}_k m,~~~\lambda_k = 1 + q^2 - 2q \cos \frac{k \pi}{N+1},~~~k=1, 2, \cdots, N.
\eeq{masses}
Rotation of the right-handed fields to the mass eigenbasis is given by 
\beq
\zero{U}_R^{j0} = \sqrt{\frac{q^2 - 1}{q^2 - q^{-2N}}} \frac{1}{q^{N-j}}, \;\; j=0,\cdots, N,  
\eeq{zeromode}
for the zero mode, and
\beq
\zero{U}_R^{jk} = \sqrt{\frac{2}{(N+1) \lambda_k}} \left[ q \sin \frac{(N-j)k \pi}{N+1} - \sin \frac{(N-j+1) k \pi}{N+1} \right],~~~~j=0, \ldots, N;~~~k = 1, \cdots N,
\eeq{UR_massive}
for the massive states. For the left-handed fields, 
\beqa
\zero{U}_L^{00} &=& 1,~~~~\zero{U}_L^{0j} = \zero{U}_L^{j0} = 0;\CR 
\zero{U}_L^{jk} &=& \sqrt{\frac{2}{N+1}} \sin \frac{jk\pi}{N+1}, \;\; j,k=1,\cdots,N.
\eeqa{UL_massive}
In terms of the pictorial representation of the clockwork in Fig.~\ref{fig:clockwork_fermion}, the massive left- and right-handed eigenmodes appear ``delocalized", mixing the fields at all sites in roughly equal measure. On the other hand, the zero mode is strongly localized. The left-handed part of the zero mode corresponds exactly to the field $\nu_L$. The right-handed zero mode consists mainly of the field $\chi_N$, with rapidly decreasing admixtures from the fields located further to the left. In particular, the contribution of $\chi_0$ is suppressed by a factor of $1/q^N$. (These features are illustrated in Fig.~\ref{fig:localization}.) When the Yukawa coupling is turned on, the resulting Dirac mass of the pseudo-zero mode is suppressed by the same factor, yielding an exponentially small neutrino mass for moderate values of $q$ and $N$. In this way, the {\it clockwork mechanism} generates a small Dirac neutrino mass without small input parameters. 

\begin{figure}[t!]
	\center
	\includegraphics[width=8cm]{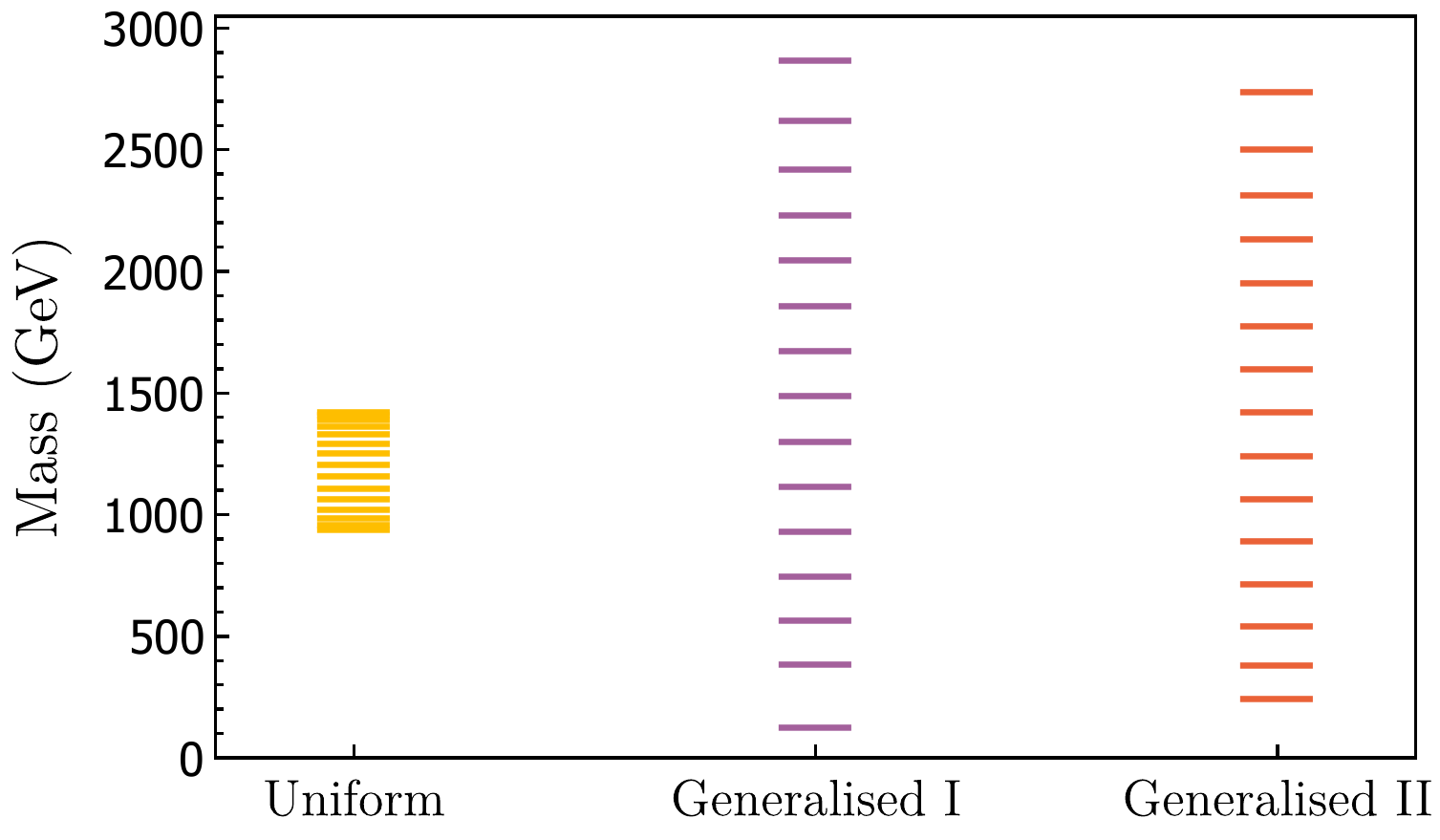}
	\caption{Spectrum of clockwork neutrinos in the uniform CW (left) and generalized Linear CW1 (center) and Linear CW2 (right) models. (For discussion of generalized CW models, see Section~\ref{sec:extended}.) In all cases, $N=15$, $m=v$, and $y=0.01$; $q=4.887$ (uniform), $0.76$ (LCW1) and $0.73$ (LCW2). The parameters were chosen so that the pseudo-zero mode neutrino mass is $m_\nu=8\cdot 10^{-2}$ eV in all cases.}
	\label{fig:spectrum}
\end{figure}

When a Yukawa coupling is present and the Higgs acquires a vev, the mass matrix has the form 
\beq
M_\alpha = m
\begin{pmatrix}
    p_\alpha & 0 & 0 & \cdots & 0 & 0 \\
	-q & 1 & 0 & \cdots & 0 & 0 \\
	0 & -q & 1 & \cdots & 0 & 0 \\
	\vdots & \vdots & \vdots & \ddots & \vdots & \vdots \\
	0 & 0 & 0 & \cdots & -q & 1
\end{pmatrix},
\eeq{mass_matrix_Y}
where $\alpha=1\ldots 3$ is the flavor index, and we defined
\beq
p_\alpha=\frac{y_\alpha v}{\sqrt{2} m}\,.
\eeq{pdef}
Here $v=246$~GeV is the Higgs vev. The spectrum consists of $N+1$ Dirac neutrinos for each flavor:
\beq
\N_j = (\N_{L\,j}, \N_{R\,j}),~~~~j=0\ldots N
\eeq{Diracs}
where 
\beq
\Psi = U_L \N_L,~~~~~X = U_R \N_R,
\eeq{states}
and we suppressed the flavor index. Since the Yukawa coupling explicitly breaks translational symmetry of the mass terms, it is no longer possible to obtain the spectrum and determine the rotation matrices $U_L$ and $U_R$ analytically. Numerical diagonalization can always be performed. However, it is also useful to obtain approximate formulas, valid in the situation when Yukawa coupling is small enough to be treated as a perturbation. The perturbation theory is developed systematically in Appendix~\ref{app:PT}. The lightest mass eigenstate is the pseudo-zero mode whose mass vanishes in the absence of the Yukawa. This state is identified with the experimentally observed (or ``active") neutrino. It has a mass (up to corrections of order $p^4$) 
\beq
m_{0,\alpha} = m \frac{p_\alpha}{q^N} \left( \frac{q^2 - 1}{q^2 - q^{-2N}} \right)^{1/2} \left( 1 - \frac{p_\alpha^2}{2(N+1)} \sum_{k=1}^N \frac{C_k}{\lambda_k} \right)\,, 
\eeq{zeromode_mass}
where $\alpha=1\ldots 3$ labels the three active neutrino mass eigenstates (this index is {\it not} summed over when repeated), and
\beq
C_k = \frac{2 q^2}{\lambda_k} \sin^2 \frac{N k \pi}{N+1}.     
\eeq{Cdef}
As expected, we obtained $m_0\sim \frac{yv}{q^N}$, allowing to generate the observed neutrino mass scale with $y\sim 1$, $q\sim$ a few, $N\sim 10$. Note that at leading order in the perturbative expansion, the active neutrino masses are independent of $m$. The remaining $N$ mass eigenstates, which we will call {\it clockwork neutrinos}, have masses (again up to corrections of order $p^4$)
\beq
m_{k,\alpha} = m \lambda_k^{1/2} \left( 1 + \,\frac{p_\alpha^2}{2(N+1)}\,\frac{C_k}{\lambda_k}\right),\,~~~k=1\ldots N. 
\eeq{ith_dude_mass}    
Clockwork neutrino masses are of order $m$, typically around the weak/TeV scale. It can be easily seen from Eq.~\leqn{masses} that the spectrum consists of $N$ states with masses in a band between $(q+1)m$ and $(q-1)m$ (up to corrections of order $p^2$); a sample spectrum is shown in Fig.~\ref{fig:spectrum}. Note that the width of the mass band is independent of $N$, so the states become more closely spaced with growing $N$: the splitting between neighboring clockwork mode masses is of order $\Delta m \sim 2m/N$. The perturbation theory developed in Appendix~\ref{app:PT} is valid only if the Yukawa shift in these masses is small compared to the splitting. This yields $p\ll 1$ as a plausible condition for validity of the perturbation theory.    

Rotations $U_L$ and $U_R$ that diagonalize the mass matrix have the form
\beq
U_{L\alpha} = \zero{U}_L \, \left( 1 + \Delta_{L\alpha} \right) ,~~~~U_{R\alpha} = \zero{U}_R\,\left( 1 + \Delta_{R\alpha} \right).
\eeq{FullUs} 
For small $p$, the rotation matrices can be computed analytically. Up to corrections of order $p^3$, we obtain
\beqa
\Delta_{R\alpha}^{i0} &=& -\Delta_{R\alpha}^{0i} = -p_\alpha^2 \sqrt{\frac{1}{N+1} \frac{q^2 - 1}{q^{2(N+1)} - 1}}  \frac{\sqrt{C_i}}{\lambda_i} \,,~~~~i=1\ldots N;\CR
\Delta_{R\alpha}^{ij} &=& \frac{p_\alpha^2}{N+1}\,\frac{\sqrt{C_i C_j}}{\lambda_j - \lambda_i}\,,~~~~i, j=1\ldots N,~~j\not=i; \CR
\Delta_{R\alpha}^{kk} &=& 0,~~~~k=0\ldots N,
\eeqa{Rs}
for the right-handed rotation, and 
\beqa
\Delta_{L\alpha}^{i0} &=& -\Delta_{L\alpha}^{0i} = p_\alpha\,\sqrt{\frac{1}{N+1} \frac{C_i}{\lambda_i}} \,,~~~~i=1\ldots N;\CR
\Delta_{L\alpha}^{ij} &=& \frac{p_\alpha^2}{N+1}\,\sqrt{\frac{\lambda_i}{\lambda_j}}\,\frac{\sqrt{C_i C_j}}{\lambda_j - \lambda_i}\,,~~~~i, j=1\ldots N,~~j\not=i; \CR
\Delta_{L\alpha}^{ii} &=& - \frac{p_\alpha^2}{N+1}\,\frac{C_i}{2\lambda_i},~~~~i=1\ldots N; \CR
\Delta_{L\alpha}^{00} &=& - \frac{p_\alpha^2}{N+1}\,\sum_{k=1}^N \frac{C_k}{2\lambda_k},
\eeqa{Ls}
for the left-handed rotation. We checked that $U_L$ and $U_R$ defined by these formulas are unitary up to terms of order $p^3$. 

Phenomenology of the clockwork neutrino sector is controlled by its contribution to weak currents. The charged current Lagrangian is
\beq
{\cal L}_{\rm CC} = g W^+_\mu J^{\mu+}_W + {\rm h.c.}, 
\eeq{Lcc} 
where $g$ is the SM weak coupling, and 
\beq
J^{\mu+}_W =\frac{V_{\alpha\beta}}{\sqrt{2}} \, \overline{e}_\alpha \gamma^\mu \nu_{L\beta} = \sum_{j=0}^{N}\frac{V_{\alpha\beta} }{\sqrt{2}} \, \bar{e}_\alpha \gamma^\mu (U_{L\beta})^{0j} P_L \N_{j \beta}. 
\eeq{Jcc}
Here $V_{\alpha\beta}$ is the standard PMNS matrix describing flavor mixing in the neutrino sector, and $P_L=\frac{1-\gamma_5}{2}$ is the left-handed projector. 
For small $p$, we obtain   
\beq
J^{\mu+}_W = \frac{V_{\alpha\beta}}{\sqrt{2}} \overline{e}_{L\alpha} \gamma^\mu P_L \left(\kappa_{0\beta} \N_{0 \beta} + \sum_{j=1}^{N} \kappa_{j\beta} \N_{j\beta} \right)\,+\,{\cal O}(p^3)\,,
\eeq{Jcc_pert}
where 
\beqa
\kappa_{0\beta} &=& 1 - \frac{p_\beta^2}{N+1}\,\sum_{k=1}^N \frac{C_k}{2\lambda_k};\CR
\kappa_{j\beta} &=& -p_\beta \,\sqrt{\frac{1}{N+1} \frac{C_j}{\lambda_j}}\,,~~~~j=1\ldots N.
\eeqa{kappas}
Physically, $\kappa_0\not= 1$ corresponds to a shift in the active neutrino charged current coupling, while $\kappa_j$ induce couplings of clockwork neutrinos to the SM electron and $W$ boson. The first effect occurs at ${\cal O}(p^2)$, while the second effect occurs at ${\cal O}(p)$. Both effects are flavor-dependent. Note that $\kappa_0^2+\sum_j \kappa_j^2=1$, as required by unitarity.    

Neutral current (NC) interactions are described by 
\beq
{\cal L}_{\rm NC} = \frac{g}{\cos\theta_w} Z_\mu J^{\mu}_Z,
\eeq{Lnc}
where $\theta_w$ is the SM Weinberg angle, and 
\beq
J^{\mu}_Z =\frac{1}{2} \overline{\nu}_{L\alpha} \gamma^\mu \nu_{L\alpha} = \frac{1}{2} \sum_{j,k=0}^{N}\overline{\N}_{j \alpha} \gamma^\mu (U_{L\alpha}^\dagger)^{j0} (U_{L\alpha})^{0k} P_L \N_{k \alpha}.
\eeq{Jnc}
For small $p$, the active-neutrino NC has the form
\beq
J^{\mu}_Z = \frac{1}{2} \overline{\N}_{0 \alpha} \gamma^\mu P_L \left(\frac{1}{2} \eta_{0\alpha} \N_{0\alpha} + \sum_{j=1}^{N} \eta_{j \alpha}  \N_{j \alpha}\right)+ {\rm h.c.}\,+\,{\cal O}(p^3),
\eeq{Jcc_pert}
where
\beqa
\eta_{0\alpha} &=& 1 - \frac{p_\alpha^2}{N+1}\,\sum_{k=1}^N \frac{C_k}{\lambda_k};\CR
\eta_{j\alpha} &=& -p_\alpha\,\sqrt{\frac{1}{N+1} \frac{C_j}{\lambda_j}}\,,~~~~j=1\ldots N.
\eeqa{etas}
Physically, $\eta_0\not= 1$ corresponds to a shift in the coupling of active neutrinos to the $Z$ boson, while $\eta_j$ terms induce off-diagonal couplings of the $Z$ to an active and a clockwork neutrino.     


\section{Experimental Constraints}
\label{sec:constraints}

The uniform clockwork model has 6 parameters: $m$, $q$, $N$, and the three Yukawa couplings $y_\alpha$. (Equivalently, Yukawa couplings can be traded for parameters $p_\alpha$ using Eq.~\leqn{pdef}.) Three combinations of these parameters correspond to active neutrino masses $m_{0,\alpha}$. Experimentally, only the two mass splittings have been measured so far: $\Delta m^2_{21}=m^2_{0,2}-m^2_{0,1}=7.2\cdot 10^{-5}$~eV$^2$ and $\Delta m^2_{32}=m^2_{0,3}-m^2_{0,2}=\pm 2.5\cdot 10^{-3}$~eV$^2$, while the overall mass scale is unknown. We will consider two possibilities: the {\it normal spectrum}, with $m_{0,1}=0$ and $m_{0,3}\gg m_{0,2}$; and the {\it degenerate spectrum}, with $m_{0,1}\approx m_{0,2}\approx m_{0,3}$ and $\sum_\alpha m_{0,\alpha}=0.2$~eV~\cite{Tanabashi:2018oca}. The degenerate spectrum corresponds to the largest values of active neutrino masses consistent with cosmology. These two choices correspond to two possible textures in the Yukawa couplings: hierarchical $y_3\gg y_2 \gg y_1$ and quasi-degenerate $y_3\sim y_2 \sim y_1$. (In the case of inverted hierarchical spectrum, $y_2\sim y_3 \gg y_1$, clockwork phenomenology is similar to the normal spectrum case.) Once the spectrum is chosen, three combinations of the six parameters are fixed. In this section, we will discuss experimental constraints on the remaining parameters. 

\subsection{Lepton Flavor Violation}

As we saw in the previous section, the clockwork model entails flavor-dependent shifts in the CC and NC couplings of SM leptons. Flavor-dependent couplings of SM leptons to massive clockwork neutrinos are also introduced. These effects induce lepton-flavor violating (LFV) processes. The tightest experimental constraint\footnote{Currently, constraints from the decay $\mu\to eee$ and the $\mu\to e$ conversion are subdominant to $\mu\to e\gamma$, but the situation may change with the next round of experiments~\cite{Hewett:2012ns,Abada:2014kba}.} on such effects is from the non-observation of the decay $\mu\to e\gamma$, whose branching ratio is currently constrained to be at most $4.2\times 10^{-13}$ at 90\%~c.l.~\cite{TheMEG:2016wtm}. 

In the clockwork model, the $\mu\to e\gamma$ branching ratio is 
\beqa
& & \mathrm{Br}(\mu \rightarrow e \gamma) = \frac{3\alpha}{8\pi}\left\lvert {\cal A}\right\rvert^2\,,\CR
& &{\cal A} = \sum_{\alpha=1}^{3} \sum_{j=0}^{N} V_{\mu\alpha} V^*_{e\alpha}  \lvert(U_{L\alpha})^{0j}\rvert^2 F\left(\frac{m_{j,\alpha}^2}{m^2_W}\right)\,,
\eeqa{mutoegamma}         
where the loop function is given by~\cite{PhysRevLett.45.1908,PhysRevD.24.1410} 
\beq
F(x) = \frac{1}{6 (1-x)^4} \left(10 - 43 x + 78 x^2 - 49 x^3 + 4 x^4 +18 x^3 \log x \right).
\eeq{F}
Within the small-$p$ perturbation expansion developed in the previous section, the first non-vanishing contribution to ${\cal A}$ occurs at order $p^2$. This contribution can be conveniently written as  
\beq
{\cal A} = \left( \frac{y^2_3 v^2}{2m^2}\right) \cdot \left[ \frac{V^*_{e3}V_{\mu3} \Delta m^2_{32}-V^*_{e1}V_{\mu1} \Delta m^2_{21}}{m_{0,3}^2}\right] \cdot \,{\cal F}(m, q, N)\,,
\eeq{arg1}
where
\beq
{\cal F}(m, q, N) = \frac{1}{N+1}\,\, \sum_{k=1}^N \frac{C_k}{\lambda_k}\, \left( F\left(\frac{m^2\lambda_k}{m^2_W}\right) - F(0) \right)\,.
\eeq{Apert}
The expression in the square brackets is ${\cal O}(0.1-1)$ depending on the assumed neutrino spectrum, while ${\cal F}\sim {\cal O}(0.1)$ for typical clockwork parameters. The experimental bound on $\mu\to e\gamma$ then roughly implies     
\beq
\frac{y_3 v}{m} \lsim 10^{-2}.
\eeq{constraint}
Either a mild hierarchy between $v$ and $m$, with clockwork states around 10 TeV, or a Yukawa coupling of order $10^{-2}$, are necessary to satisfy this bound. In either case, the bound does not invalidate the original motivation for the clockwork model, since small parameters of the required size are by no means unusual in the SM. The second case, $m\sim v$ and $y\sim 10^{-2}$, is especially interesting from the phenomenological point of view, since the clockwork states are light enough to be produced at the LHC and the proposed lepton colliders. We will consider their collider phenomenology in Section~\ref{sec:coll}.    

\begin{figure}[t!]
	\center
	\includegraphics[width=7.5cm]{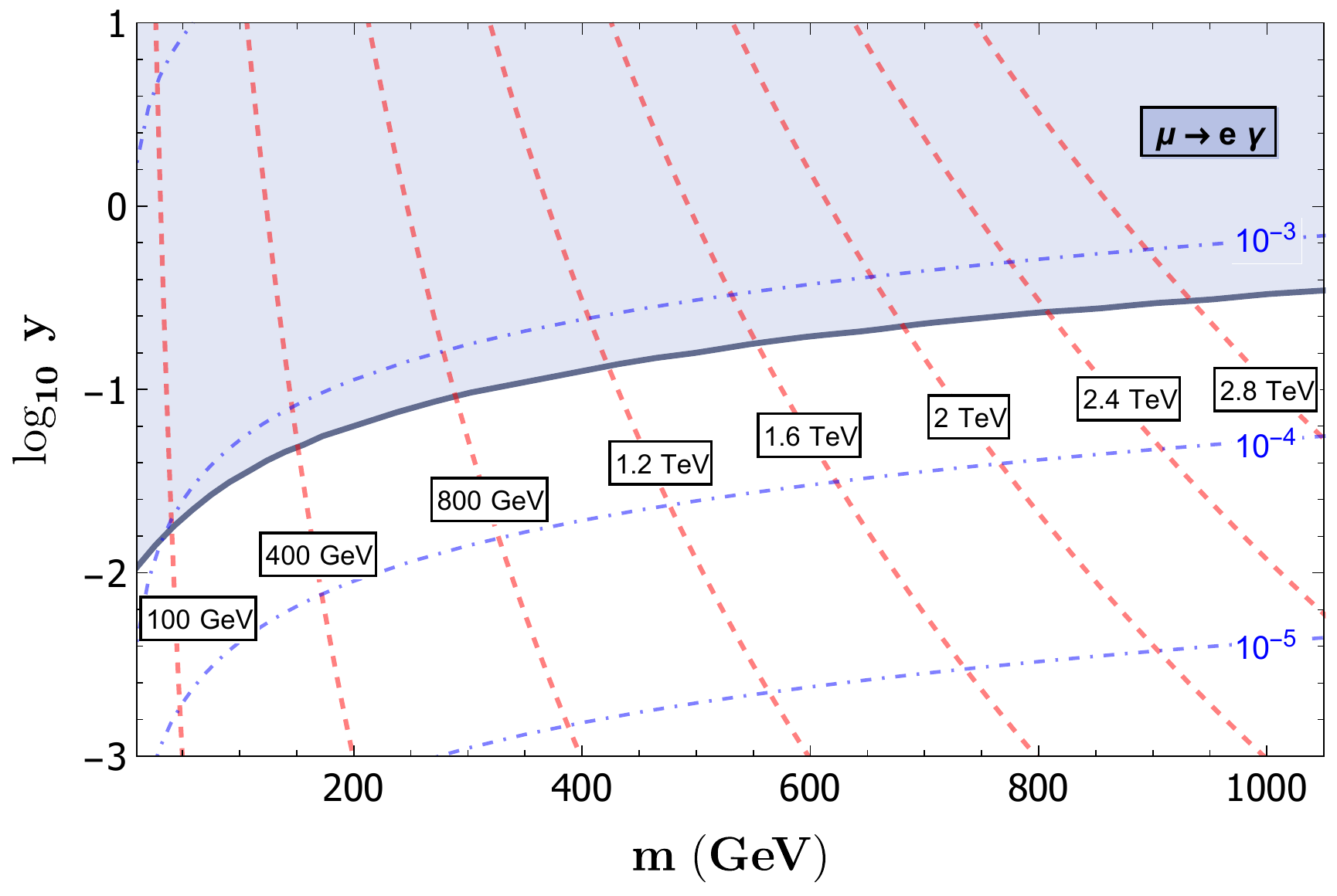}
	\includegraphics[width=7.5cm]{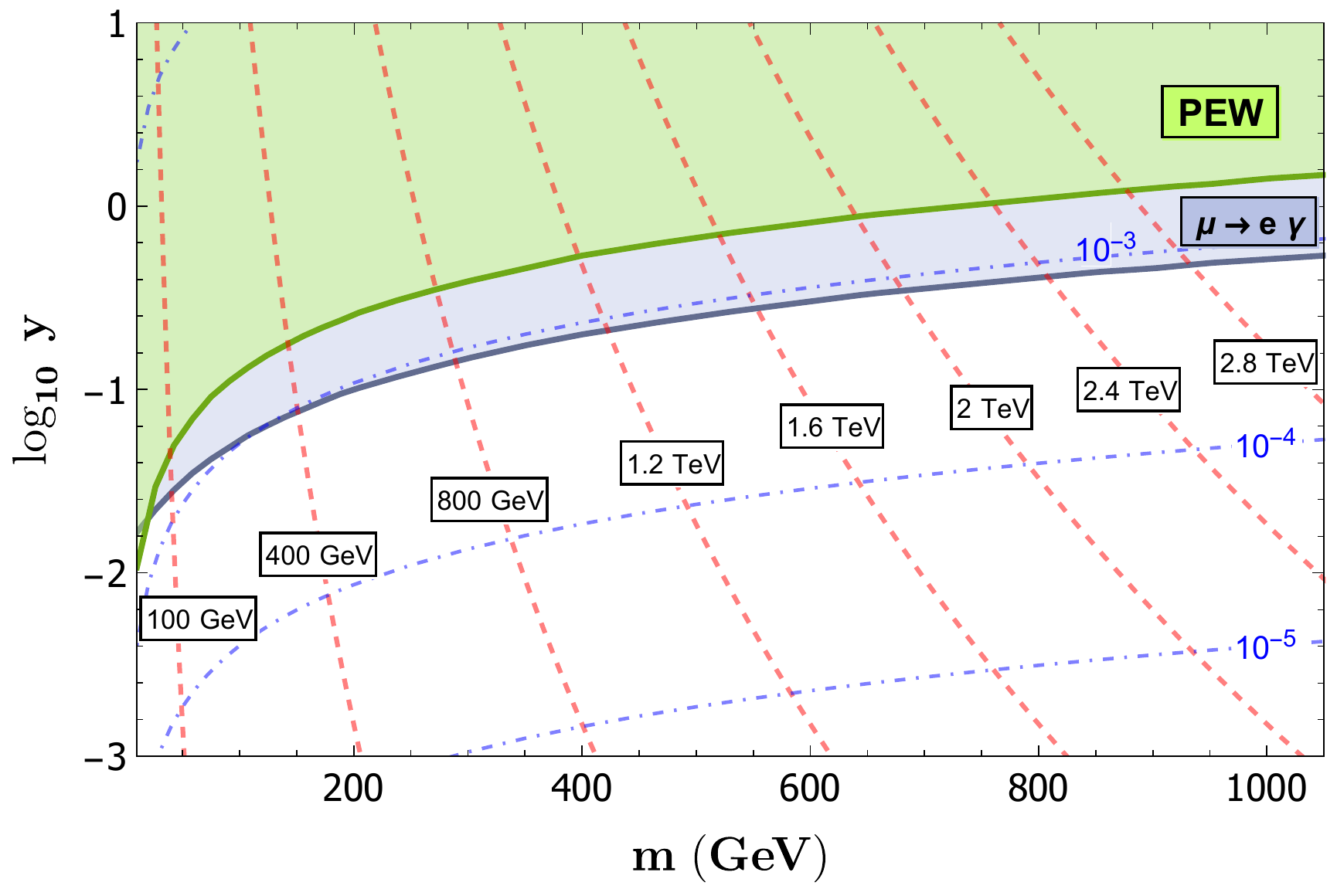}
	\caption{Constraints on the parameter space of the uniform clockwork model from $\mu\to e\gamma$ and precision electroweak fits. Left panel: normal (hierarchical) spectrum of active neutrinos. Right panel: degenerate spectrum of active neutrinos. Dashed/red lines indicate the mass of the lightest clockwork neutrino, while dot-dash/blue lines indicate the coupling of this state to the SM gauge currents.}
	\label{fig:constraints_uniform}
\end{figure}

Implications of non-observation of $\mu\to e\gamma$ for the model parameter space are illustrated by Fig.~\ref{fig:constraints_uniform}. In these plots, we fix the active neutrino spectrum (as discussed at the beginning of this section), and choose $N=20$. We choose the clockwork mass $m$ and the Yukawa coupling $y_3$ as the remaining two degrees of freedom to describe the model parameter space, and present the constraints in terms of these parameters. In agreement with the intuition from Eq.~\leqn{constraint}, we observe that clockwork neutrinos at the weak scale, ${\cal O}(100)$ GeV, can be consistent with the $\mu\to e\gamma$ constraint for moderately small Yukawas, $y_3\sim 10^{-2}$.      

The $\mu\to e\gamma$ rate can also be computed without resorting to small-$p$ perurbation theory, by diagonalizing the mass matrix numerically and using Eq.~\leqn{mutoegamma}. We find that the constraints derived using this procedure are in excellent agreement with the results of a perturbative analysis. More generally, we find that the small-$p$ perturbation theory works well throughout the part of the parameter space allowed by the $\mu\to e\gamma$ constraint.      

\subsection{Precision Electroweak Constraints}

In the clockwork model, couplings of the active neutrinos to the SM gauge currents are shifted away from the SM values. This effect is described by shifts of $\kappa_0$ and $\eta_0$ parameters away from 1, see Eqs.~\leqn{kappas} and~\leqn{etas}. Such shifts affect precision electroweak (PEW) fits: for example, $\kappa_0\not=1$ modifies the lifetime of the muon, while $\eta_0\not=1$ modifies the invisible width of the $Z$ boson. In general, these shifts are flavor-dependent. In the case of normal active neutrino spectrum, the flavor-dependence is of the same order as the overall effect; in this situation, we expect that the LFV constraints such as $\mu\to e\gamma$ are much stronger than the flavor-diagonal PEW constraints. On the other hand, in the case of degenerate spectrum, the flavor-dependence in $\kappa_0$ and $\eta_0$ is small compared to their overall size. In this case, it is not {\it a priori} obvious whether LFV or PEW constraints would dominate.   

To derive the PEW constraint, we used the three best-measured PEW observables ($m_Z$, $\alpha$ and $\Gamma_\mu$) as inputs to fix the underlying SM parameters ($g$, $g^\prime$ and $v$), and performed a $\chi^2$ fit to the other PEW observables listed in Ref.~\cite{Tanabashi:2018oca}. Note that a shift in charged-current coupling $\kappa_0$ affects the relation between $\Gamma_\mu$ and $v$; this effect was consistently taken into account in the fit. The 95\%~c.l bound on the clockwork parameter space imposed by the PEW fit is shown in Fig.~\ref{fig:constraints_uniform}. We conclude that even in the degenerate spectrum case, the LFV bounds on the model parameters are currently stronger than the PEW constraint.    

\section{Generalized Clockwork Neutrinos}
\label{sec:extended}

The uniform clockwork model, proposed in Ref.~\cite{Giudice:2016yja} and developed in detail in Section~\ref{sec:base}, is only one representative of a much broader class of clockwork models that provide an exponentially small Dirac neutrino mass. As a more general example, consider a model with the same set of clockwork fields with diagonal and nearest-neighbor mass terms as before, but allow the nearest-neighbor (link) mass to vary along the clockwork chain. Using the same notation as in Section~\ref{sec:base}, the mass matrix for each neutrino flavor $\alpha$ is given by   
\beq
M_\alpha = m
\begin{pmatrix}
	p_\alpha & 0 & 0 & \cdots & 0 & 0 \\
	-q_1 & 1 & 0 & \cdots & 0 & 0 \\
	0 & -q_2 & 1 & \cdots & 0 & 0 \\
	\vdots & \vdots & \vdots & \ddots & \vdots & \vdots \\
	0 & 0 & 0 & \cdots & -q_N & 1
\end{pmatrix},
\eeq{mass_matrix_Y}
where $q_i$ are dimensionless parameters. If the Yukawa coupling is turned off, $p_\alpha=0$, this mass matrix has zero determinant, and there is a massless zero-mode. The left-handed component of the zero mode is identical to $\nu_L$. The right-handed component is a linear combination of the clockwork fields:
\beq
{\cal N}_{R0}=\sum_{i=0}^N v_i \chi_i,
\eeq{RHmode}
where $v$ is the eigenvector of $M$ corresponding to the zero eigenvalue:
\beq
\begin{pmatrix}
0 & 0 & 0 & & 0 \\
-q_1   & 1     &         &        &    \\
0      & -q_2  & 1       &        &    \\
\vdots &       &         & \ddots &    \\
&       &         & -q_N   & 1
\end{pmatrix}
\begin{pmatrix}
v_0 \\
v_1 \\
v_2 \\
\vdots \\
v_N
\end{pmatrix} = 0.
\eeq{eigeneq}
Solving these linear equations iteratively yields
\beq
v_i = v_0 \prod_{j=1}^i q_j\,.
\eeq{solution} 
The element $v_0$ is unconstrained by the eigenvalue problem, but is fixed by the normalization condition $v^T v = 1$, which yields
\beq
v_0 = \frac{1}{\sqrt{1+q_1^2+(q_1 q_2)^2+\cdots+(q_1 q_2\cdots q_N)^2}} < \frac{1}{q_1 q_2\cdots q_N}.
\eeq{v0}
As long as all (or most of) $q_i$'s are larger than one, the admixture of the field $\chi_0$ in the right-handed zero-mode is suppressed ``exponentially" ({\it i.e.} by the product of $q_i$'s). When the Yukawa is turned on, the zero-mode acquires an exponentially suppressed mass: at leading order in the small-$p$ expansion, 
\beq
m_{0, \alpha} = m p_\alpha v_0 < \frac{y_\alpha v}{\sqrt{2} q_1 q_2\cdots q_N}.
\eeq{zeromass}
The observed hierarchy between the weak scale and the neutrino masses can be generated, without introducing small or large parameters, for a broad variety of $\{q_i\}$ choices. 

\begin{figure}[t!]
	\center
	\includegraphics[width=7.5cm]{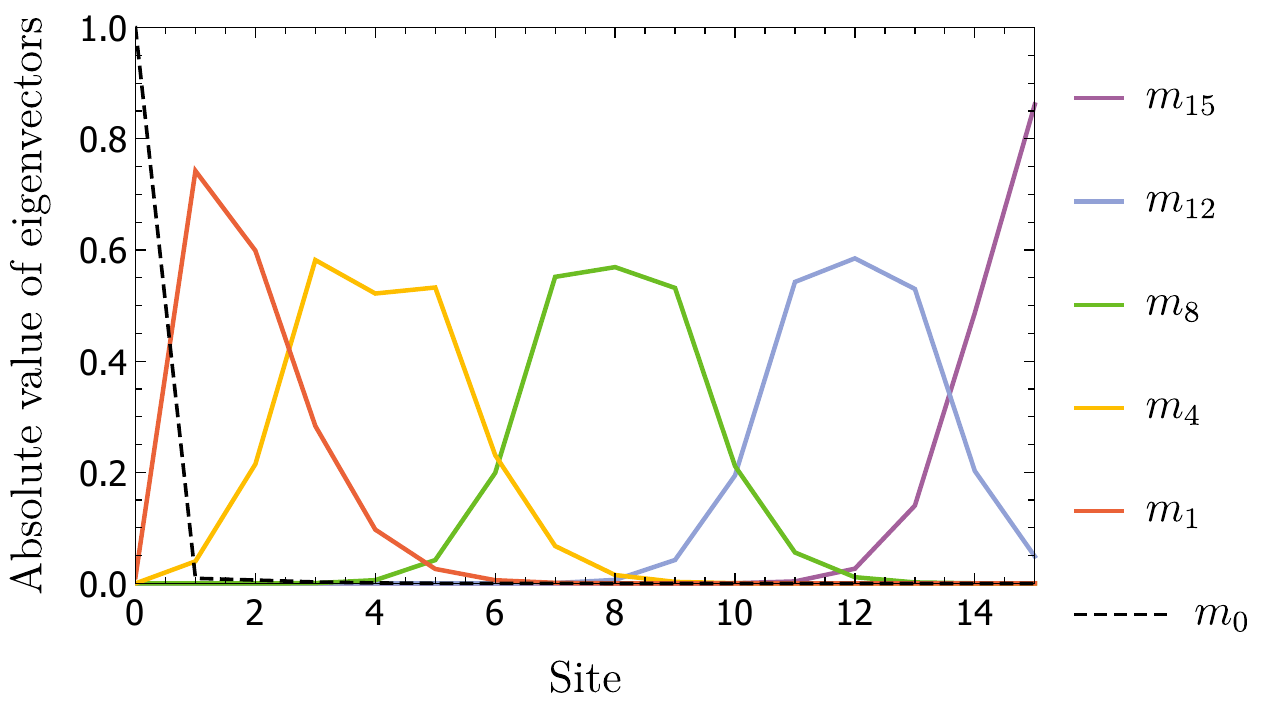}
	\includegraphics[width=7.5cm]{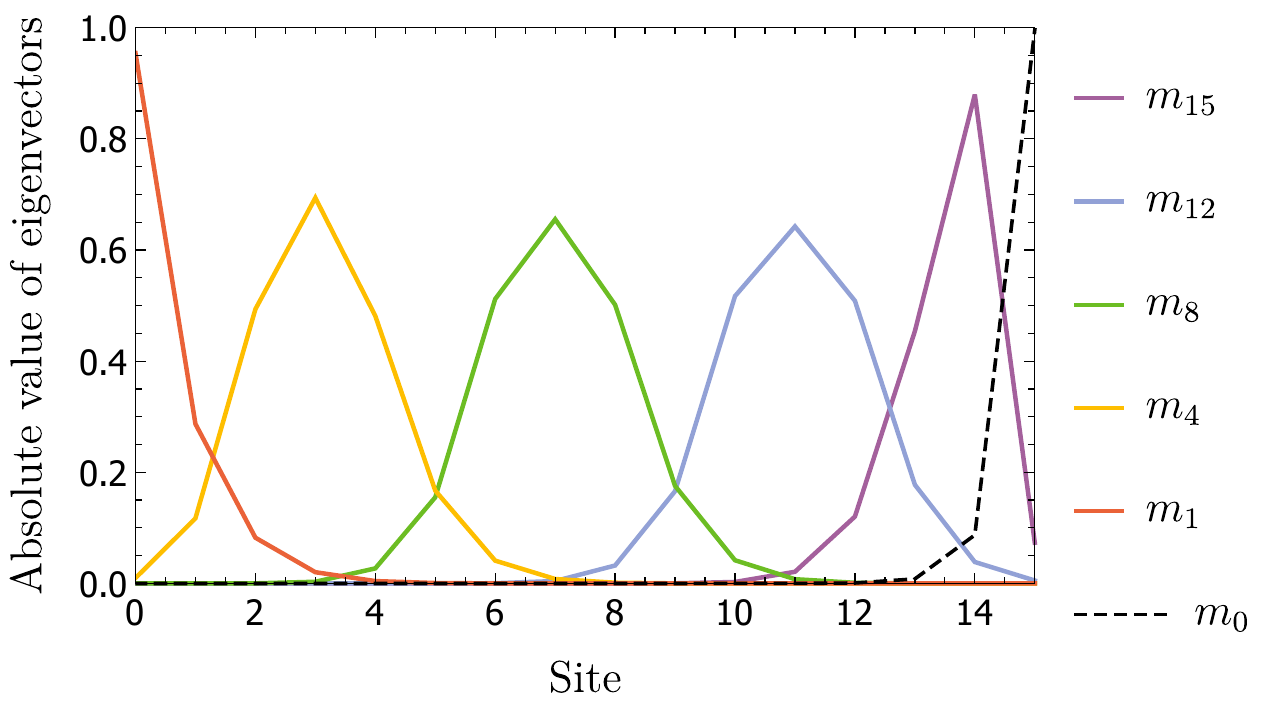}
	\includegraphics[width=7.5cm]{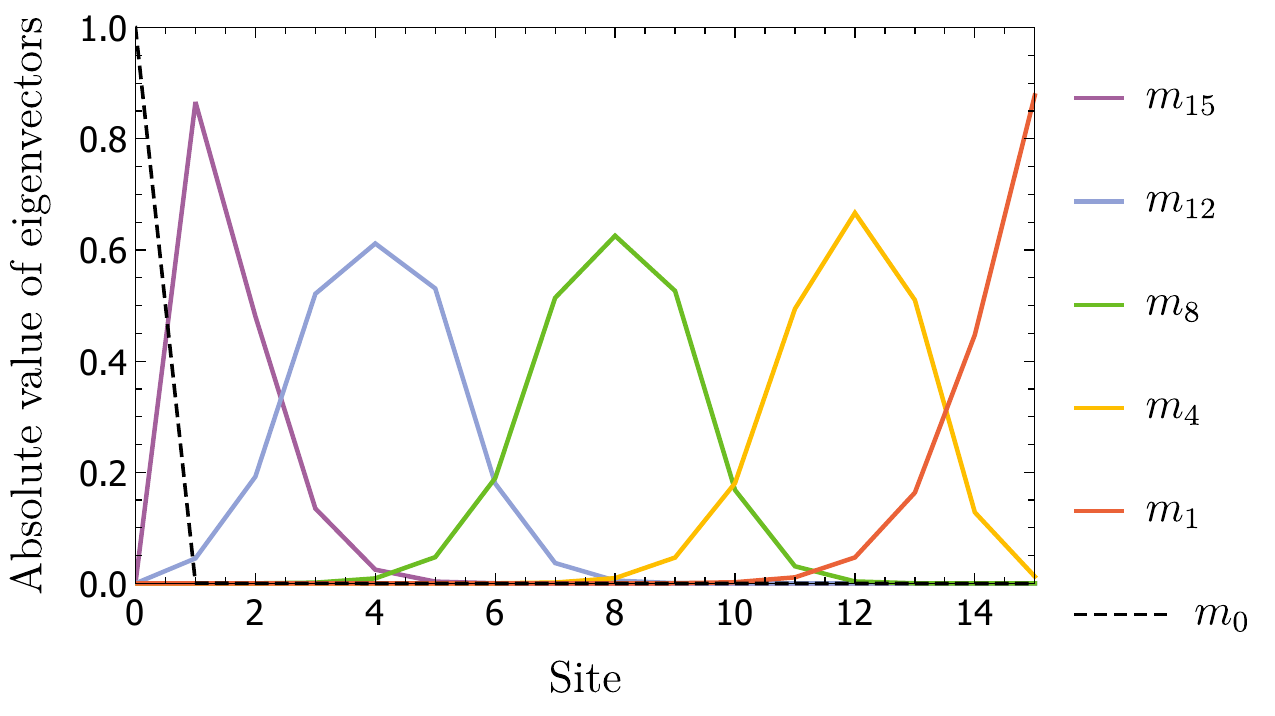}
    \includegraphics[width=7.5cm]{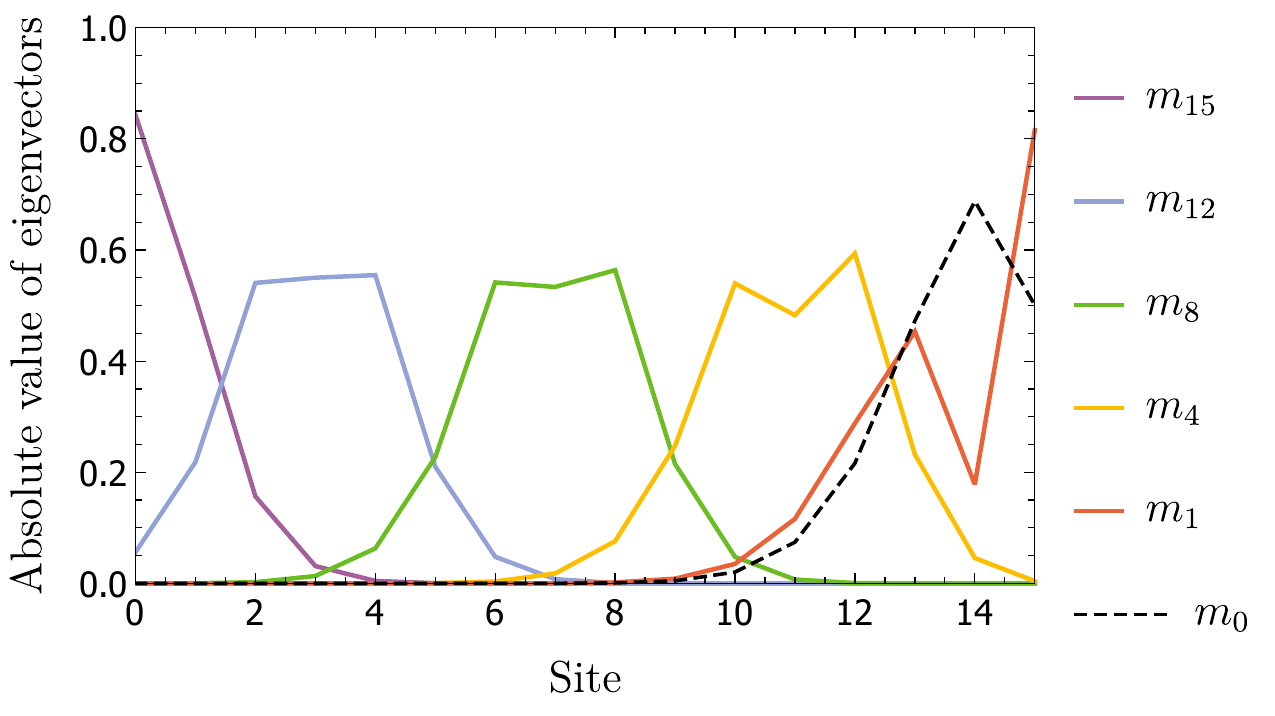}
   	\caption{Composition of the left-handed (left panel) and right-handed (right panel) mass eigenmodes in terms of the original clockwork fields in the Linear CW 1 model (top line) and Linear CW 2 model (bottom line). In both cases, $N=15$, $m=v$, and $y=0.01$; $q=0.76$ in the LCW1 model and $0.73$ in the LCW2 model.}
	\label{fig:gen_profiles}
\end{figure}

Smallness of the right-handed zero mode at the SM site is the common feature of all such models. However, the properties of excited clockwork states can vary drastically depending on the model. In particular, we observed that in the uniform model, the clockwork spectrum consists of a band of states, centered at the clockwork scale $m$ and separated by $\sim mq/N$. These clockwork states are delocalized, mixing fields at all sites in roughly equal measure. In other clockwork models, these features may be quite different. As a concrete example, consider two {\it Linear Clockwork} models:   
\beqa
{\rm Linear~CW~1:}~~&~&~q_i = qi,~i=1\ldots N; \CR
{\rm Linear~CW~2:}~~&~&~q_i = q(N+1-i),~i=1\ldots N; 
\eeqa{linearCW}
where $q>1$ is no longer required as long as $\Pi q_i \gg 1$. Sample spectra of CW neutrino modes in these two models are shown in Fig.~\ref{fig:spectrum}; unlike the uniform model, the states are no longer confined to a relatively narrow mass gap, but instead are spread out in mass similar to traditional Kaluza-Klein theories (though unlike KK theories, the number of modes is finite). The composition of the right-handed components of mass eigenmodes in terms of the original clockwork fields in these models is illustrated in Fig.~\ref{fig:gen_profiles}. In both cases, the zero mode is exponentially suppressed at the SM (leftmost) site, as expected. Contrary to the Uniform CW, each excited mode is to a good approximation localized at a single site. In the LCW1 model, the lightest clockwork mode is localized on the SM site, while in LCW2, the heaviest  clockwork mode is localized on the SM site. These modes dominate the phenomenology, since couplings of all other modes to the SM are strongly suppressed.     

\begin{figure}[t!]
	\center
	\includegraphics[width=7.5cm]{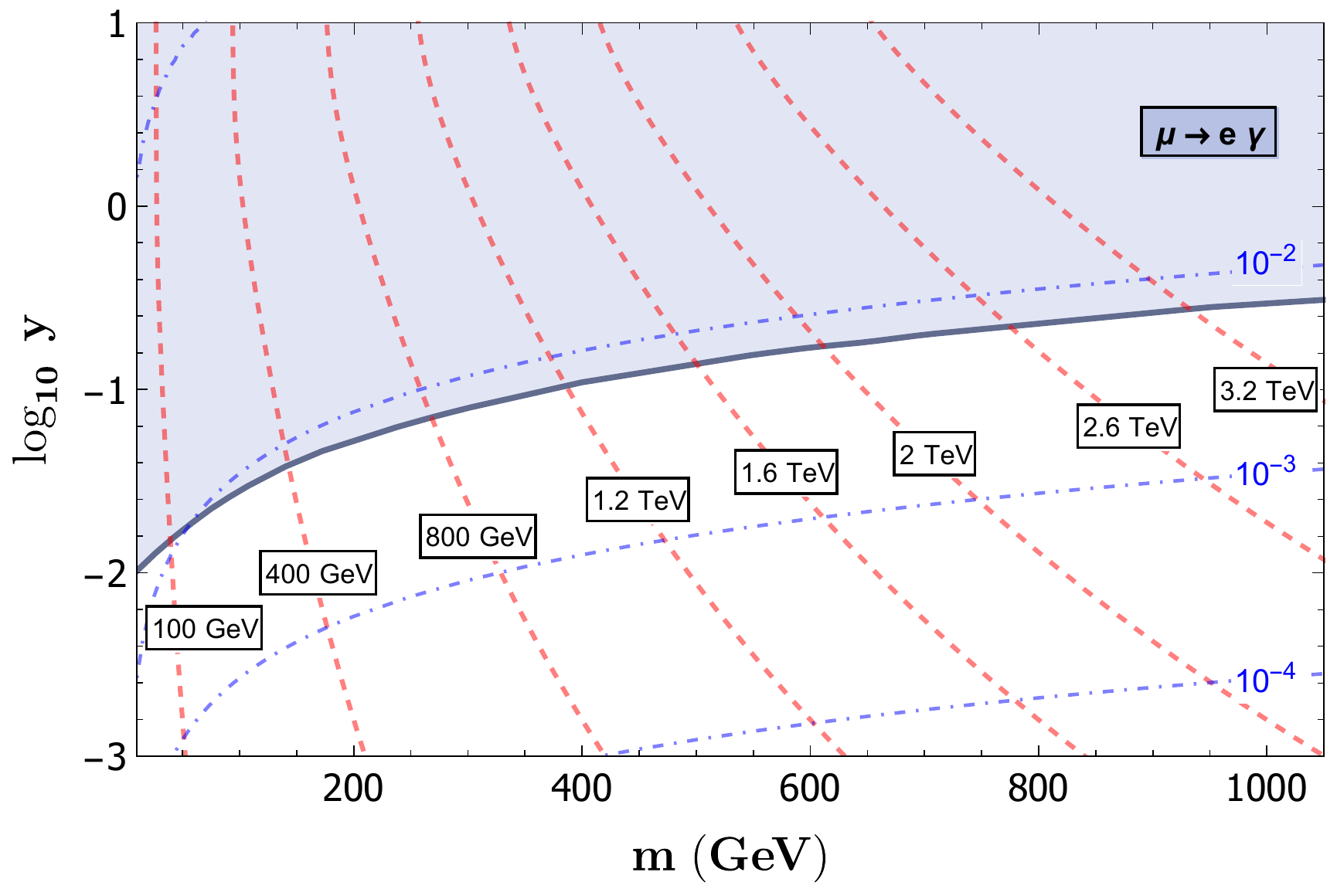}
	\includegraphics[width=7.5cm]{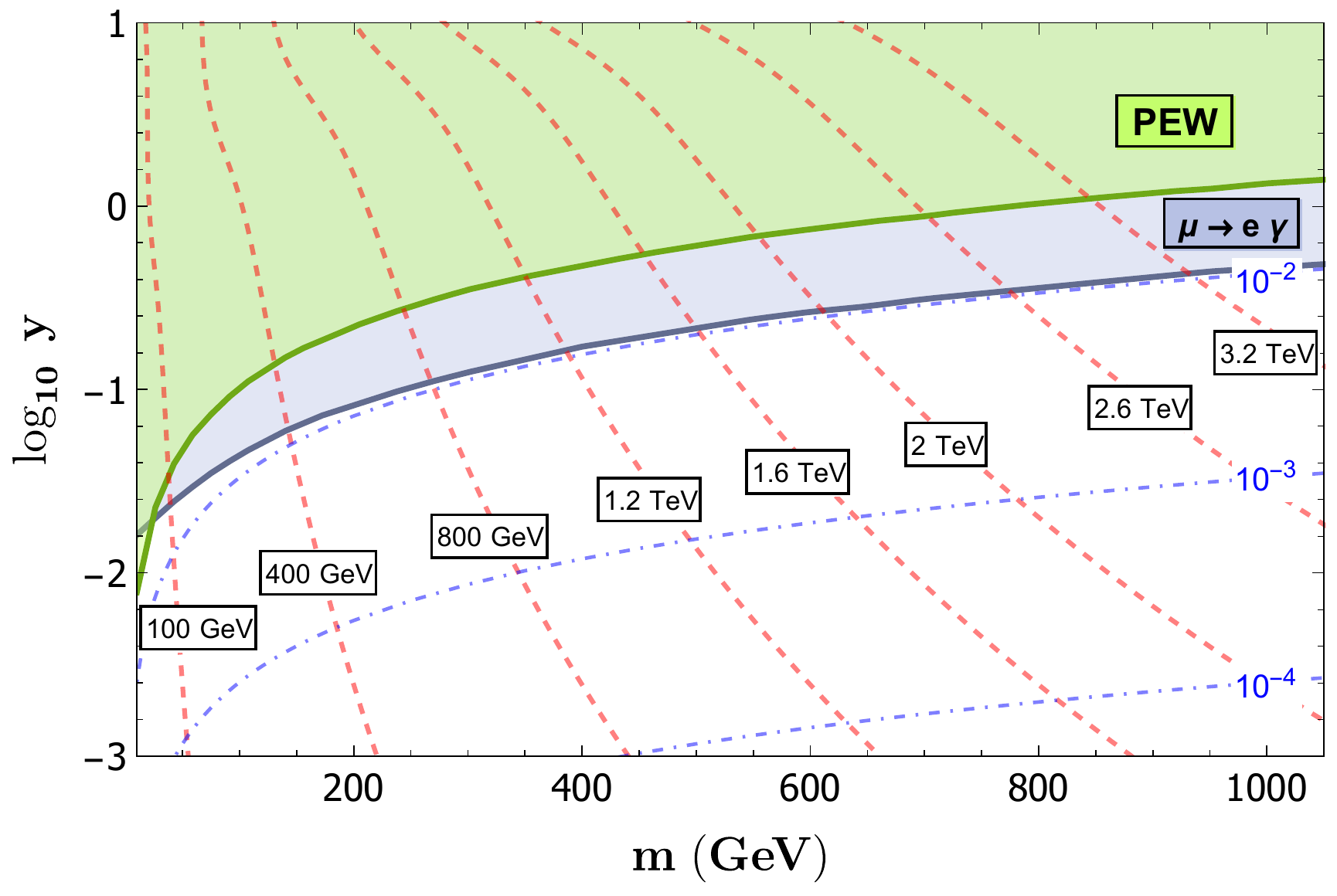}
	\includegraphics[width=7.5cm]{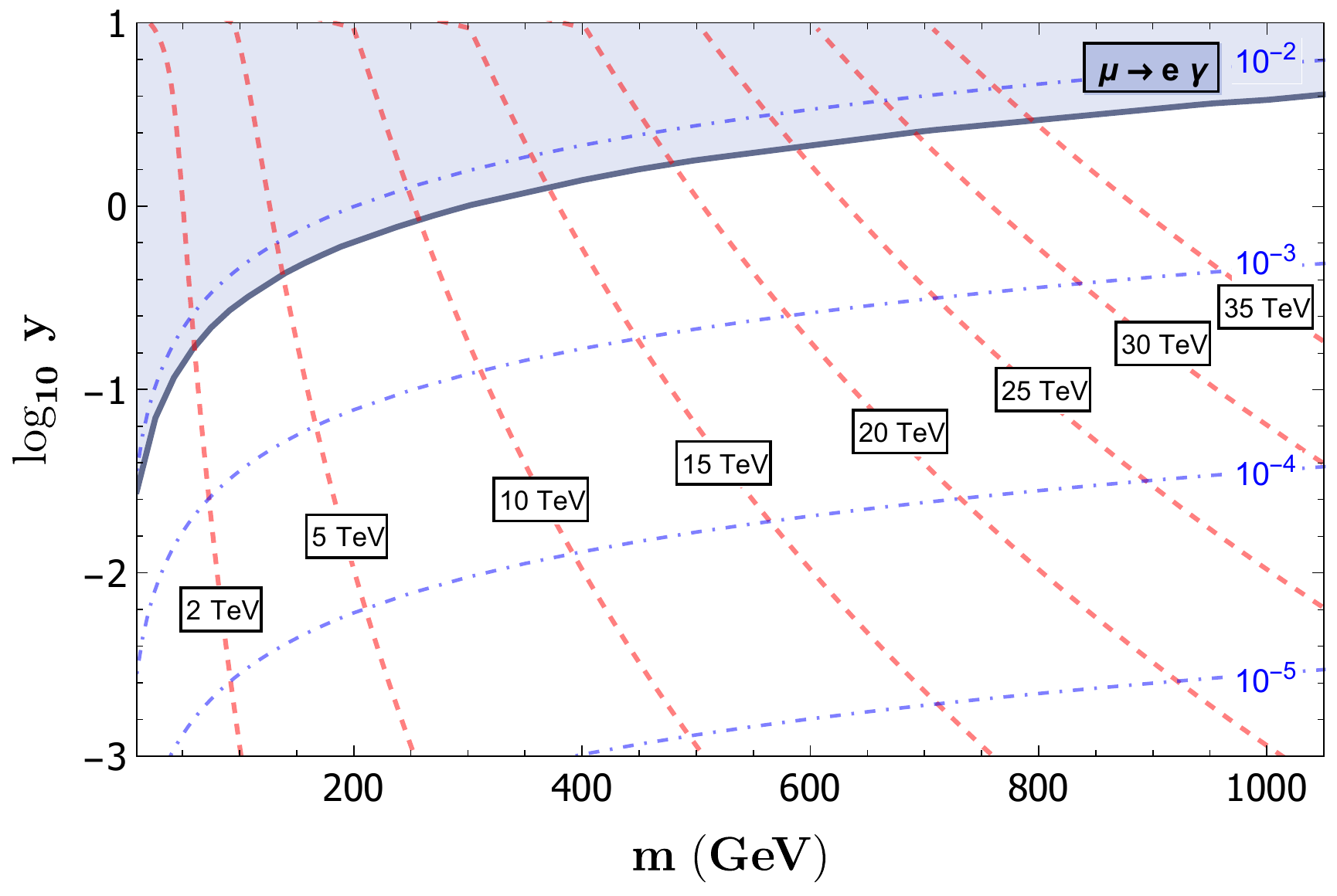}
    \includegraphics[width=7.5cm]{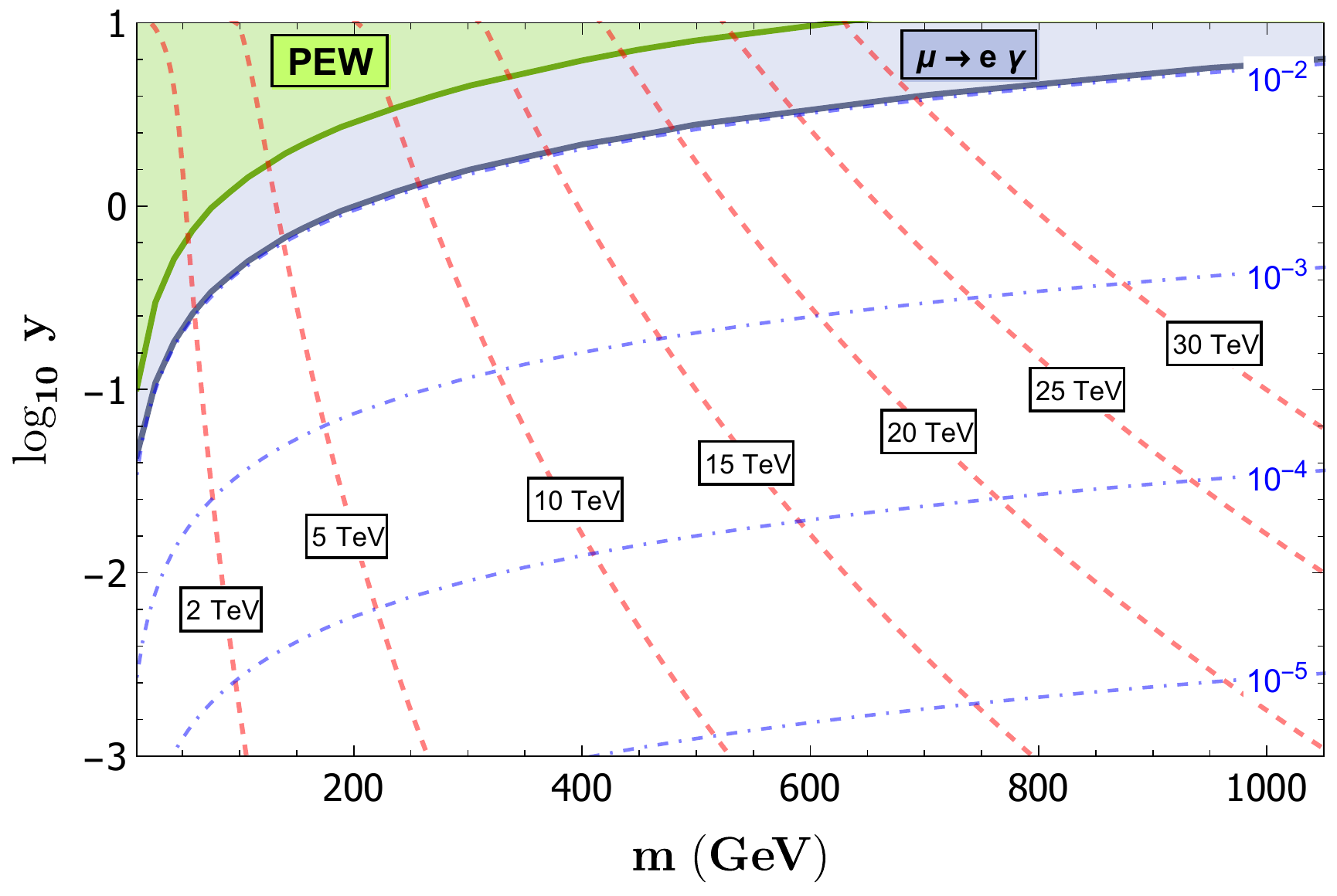}
	\caption{Top row: Constraints on the parameter space of the Linear CW 1 model from $\mu\to e\gamma$ and precision electroweak fits. Left panel: normal (hierarchical) spectrum of active neutrinos. Right panel: degenerate spectrum of active neutrinos. Dashed/red lines indicate the mass of the lightest clockwork neutrino, while dot-dash/blue lines indicate the coupling of this state to the SM gauge currents. Bottom row: same, for the Linear CW 2 model. In the case of Linear CW 2 model, the masses and couplings of the heaviest clockwork state are plotted, since other clockwork states have strongly suppressed interactions with the SM.}
	\label{fig:constraints_LCW}
\end{figure}

Experimental constraints on the Linear CW models are shown in Fig.~\ref{fig:constraints_LCW}. In both cases, we fixed $N=10$, while three more parameters are fixed by the choice of the active neutrino mass spectrum, as in the uniform model case. In the LCW1 model, constraints on the lightest clockwork state are somewhat weaker than in the uniform model: for the same mass, the allowed coupling of this state to the SM gauge currents is about one order of magnitude stronger in the LCW1 compared to the uniform model. The main reason for this is that in the LCW1 model, the lightest state alone dominates the constraints, while in the uniform model, all clockwork states give comparable contributions, yielding a stronger constraint on each one. In the LCW2 model, the state that has a significant coupling to the SM is at the top of the clockwork spectrum, and it tends to be quite heavy (in a few-TeV range) for models consistent with experimental constraints. As a result, LCW2 model is not an interesting target for TeV-scale collider phenomenology.        

\section{Collider Phenomenology}
\label{sec:coll}

We have established that current flavor and PEW constraints do not preclude the possibility that the clockwork neutrino states are within kinematic reach of the LHC and future colliders currently under discussion. In this section, we will study the associated phenomenology. 

\begin{table}[t!]
	\begin{center}
	\begin{tabular}{|c||r|r|r|r||r|r|r|}
		\hline
		Name & $N$ & $q$ & $m$ (GeV) & $y$  & $m_1$ (GeV) & $g_1$ & $m_N$ (GeV)  \\ \hline
		U100 & 20 & 3.45 &  40 & [0.0207, 0.0209, 0.0269] &  98.6  & 0.0014 & 177.7   \\
		U400 & 16 & 5.00 & 100 & [0.0541, 0.0546, 0.0704] & 402.1  & 0.0011 & 598.6  \\
		U750 & 17 & 4.70 & 200 & [0.0947, 0.0956, 0.1232] & 743.9  & 0.0010 & 1137.5  \\
		U1000 & 18 & 4.40 & 310 & [0.1362, 0.1376, 0.1773] & 1059.5 & 0.0009 & 1670.6  \\ \hline
	\end{tabular}
	\caption{Benchmark points (BPs) for the uniform clockwork model. First column: BP name. Next 4 columns: model input parameters. Last 3 columns: mass of the lightest CW neutrino; its coupling to weak current; and mass of the heaviest CW neutrino.}
	\label{tab:UBPs}
\end{center}
\end{table}

\begin{table}[t]
	\begin{center}
	\begin{tabular}{ |c||r|r|r|r||r|r|r| }
		\hline
		Name & $N$ & $q$ & $m$ (GeV) & $y$  & $m_1$ (GeV) & $g_1$ & $m_N$ (GeV)   \\ \hline
		G100 & 10 & 2.60 & 40  & [0.0178, 0.0180, 0.0232] & 101.6 & 0.0182 & 1044.7   \\
		G300& 11 & 2.05 & 160 & [0.0373, 0.0377, 0.0485] & 315.6 & 0.0122 & 3631.2   \\
		G750& 11 & 2.20 & 360 & [0.0811, 0.0819, 0.1055] & 765.9 & 0.0109 & 8760.9   \\ \hline
	\end{tabular}
	\caption{Benchmark points (BPs) for the generalised clockwork model (LCW1). First column: BP name. Next 4 columns: model input parameters. Last 3 columns: mass of the lightest CW neutrino; its coupling to weak current; and mass of the heaviest CW neutrino.}
	\label{tab:GBPs}
	\end{center}
\end{table}

For the collider study, we selected 4 benchmark points (BPs) in the uniform clockwork model, listed in Table~\ref{tab:UBPs}, and 3 BPs in the generalized model LCW1, listed in Table~\ref{tab:GBPs}. The lightest clockwork states at these BPs span the range between 100 GeV and 1 TeV, making them realistic targets for the current and near-future colliders. All BPs are allowed by the existing LFV and precision electroweak constraints. 

\begin{figure}[t]
	\center
	\includegraphics[width=7cm]{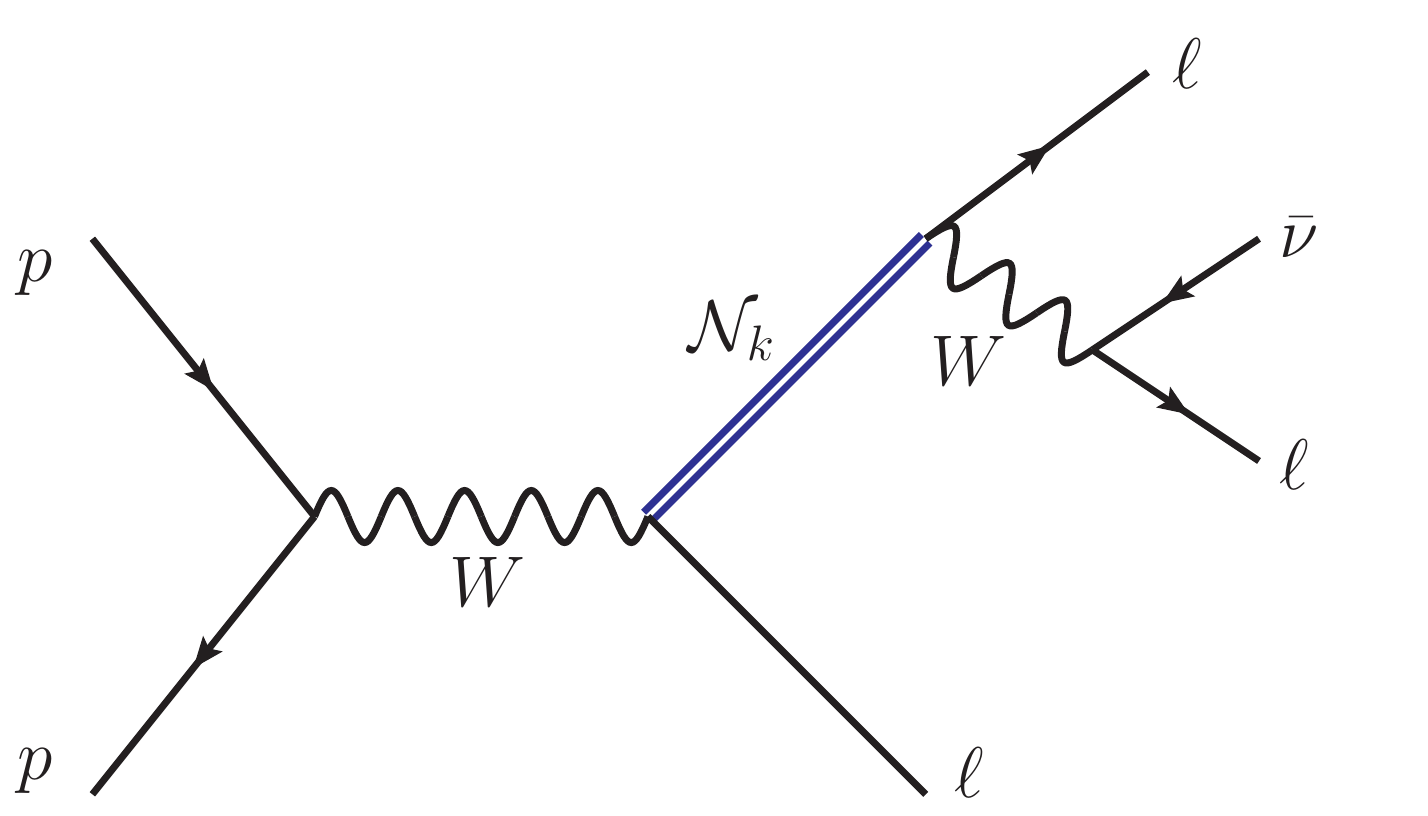}
	\includegraphics[width=7cm]{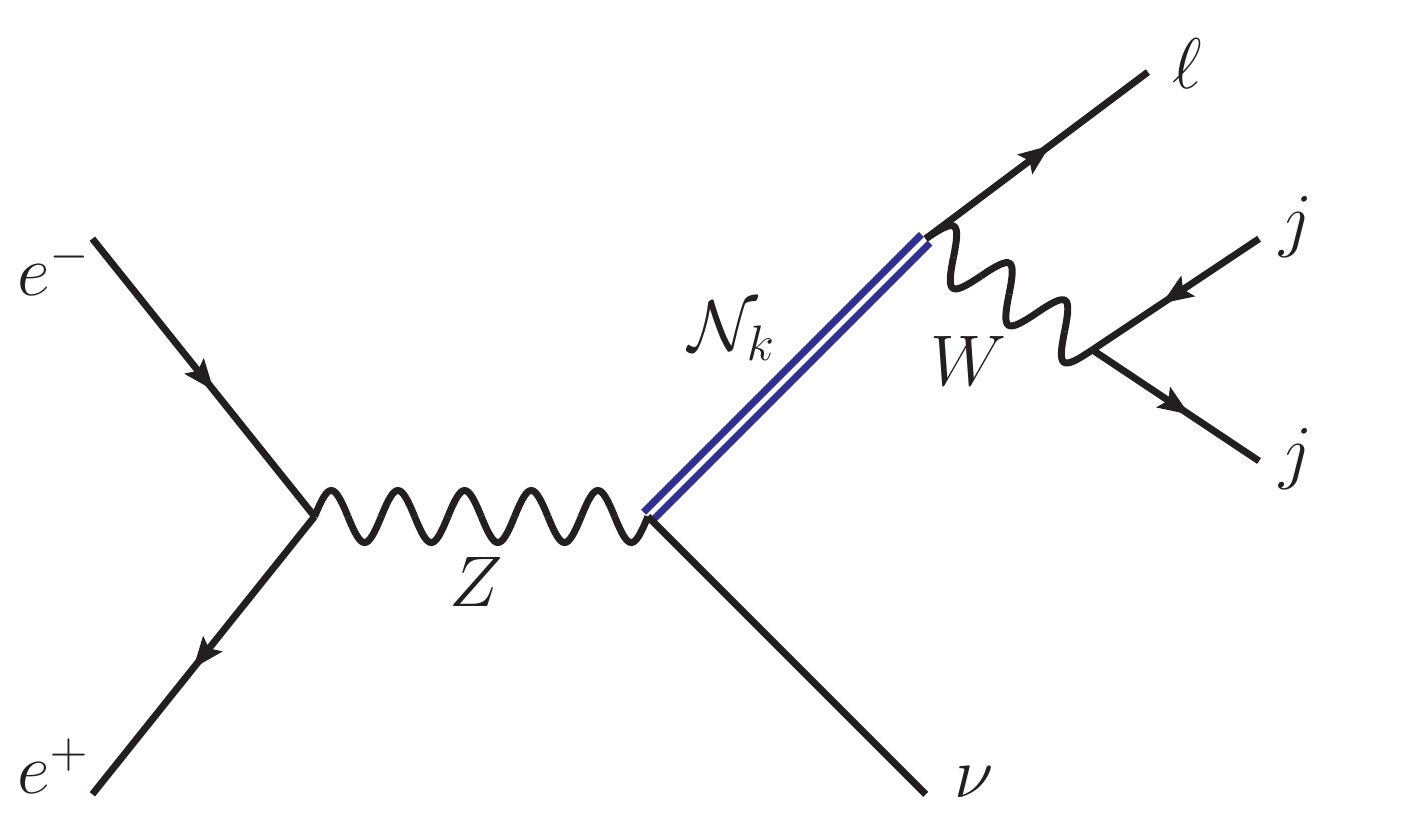}
	\caption{Production of heavy clockwork modes $\mathcal{N}_k$ at hadron (left) and lepton (right) colliders.}
	\label{fig:collider_signal_process}
\end{figure}

The simplest processes for heavy clockwork neutrino production involve $s$-channel exchange of electroweak gauge bosons, illustrated in Fig.~\ref{fig:collider_signal_process}. For hadron colliders, we focus on the $W^*$ exchange process, since an additional charged lepton in the final state improves observability of the signal. Once produced, $\mathcal{N}_k$ states promptly decay, with the dominant decay modes $\ell W$ and ${\cal N}_0 Z$. (For $k>1$, cascade decays involving intermediate CW modes may be kinematically allowed, but the corresponding couplings are sub-dominant in $p$ expansion, and branching ratios are small.) For our study, we focus on the charged-current decay $\mathcal{N}\to \ell W$. The uniform and generalized (LCW1) clockwork models were implemented in {\tt FeynRules}~\cite{Alloul:2013bka,Degrande:2011ua}. The signal and relevant backgrounds were simulated using \texttt{MadGraph@aMC}~\cite{Alwall:2011uj,Alwall:2014hca}. The parton-level events are passed to \texttt{Pythia8}~\cite{Sjostrand:2007gs} for hadronization and then to \texttt{Delphes3}~\cite{deFavereau:2013fsa} to incorporate detector effects and jet reconstruction. 

\begin{table}[t]
	\begin{center}
		\begin{tabular}{ |r||c|c||c|c|c| }
			\hline
			\multirow{2}{*}{BP} & \multicolumn{2}{c||}{$pp\to 3\ell + \met$} & \multicolumn{3}{c|}{$e^+e^-\to \ell\nu jj$} \\
		\cline{2-6}
			      & 14~TeV                & 100~TeV               & 250~GeV & 500~GeV & 3~TeV \\
			\hline
			U100  & 0.66                & 4.2                  & 7.4   & 12.8  &  3.9  \\
			G100  & 1.40                  & 8.6                & 4.3   & 7.4    &  1.34  \\
			U400  & 3.0 $\times 10^{-3}$ & 0.032               & --      & 0.81   &  7.6  \\
			G300  & 0.014                & 0.12                & --      & 5.8   &  6.1 \\
			U750  & 2.6 $\times 10^{-4}$ & 5.0 $\times 10^{-3}$ & --      & --      &  5.9 \\
			G750  & 5.0 $\times 10^{-4}$ & 8.0 $\times 10^{-3}$ & --      & --      &  7.9 \\
			U1000 & 5.0 $\times 10^{-5}$ & 1.7 $\times 10^{-3}$ & --      & --      &  2.3  \\
			\hline			
		\end{tabular}
		\caption{Cross sections of CW neutrino signatures at hadron and lepton colliders, before selection cuts (in fb). Acceptance cuts have been applied at the parton level: $\Delta R\geq 0.4$ for all visible object pairs; $p_T(\ell)\geq 10 (20)$~GeV for hadron (lepton) colliders; $p_T(j)\geq 20$ GeV; $|\eta_j| < 5$; $|\eta_\ell|<2.5$.}
		\label{tab:Xsections}
	\end{center}
\end{table}

For hadron colliders, the signatures of CW neutrino production are $3\ell + \met$ and $2\ell + 2j$, corresponding to leptonic and hadronic $W$ decays respectively.
(Here $\ell=e$ or $\mu$; we do not include taus in the analysis.) The trilepton signature has a slightly smaller rate but significantly lower backgrounds. Signal cross sections in this channel, before selection cuts, are listed in Table~\ref{tab:Xsections}. These are total cross sections, summed over CW neutrino flavor and mode number $k$. Note that the structure of CW neutrino couplings ensures that two of the leptons form a same-flavor, opposite-charge (SFOC) pair, while the third lepton (from $W$ decay) may be the same or opposite flavor. The main irreducible background to this search is $pp\to WZ$. LHC experiments have published searches in the $3\ell + \met$ channel based on 35 fb$^{-1}$ integrated luminosity at $\sqrt{s}=13$ TeV. Benchmark points U100 and G100 predict a few tens of signal events in this sample, before selection cuts. For all other benchmark points, cross sections are too small to get an appreciable number of events. CMS collaboration's search for sterile neutrino~\cite{Sirunyan:2018mtv} was optimized for a signature similar to our CW neutrino, and is expected to have the best sensitivity. We have recast this search to estimate the limits on the CW neutrino model. We find that the benchmark points U100 and G100 are not ruled out by this search. Since U100 and G100 points provide nearly-maximal LHC signals consistent with LFV and PEW constraints in their respective models, we conclude that the current LHC constraints are not yet competitive with those discussed in Section~\ref{sec:constraints}.  

\begin{figure}[t]
	\center
	\includegraphics[width=7cm]{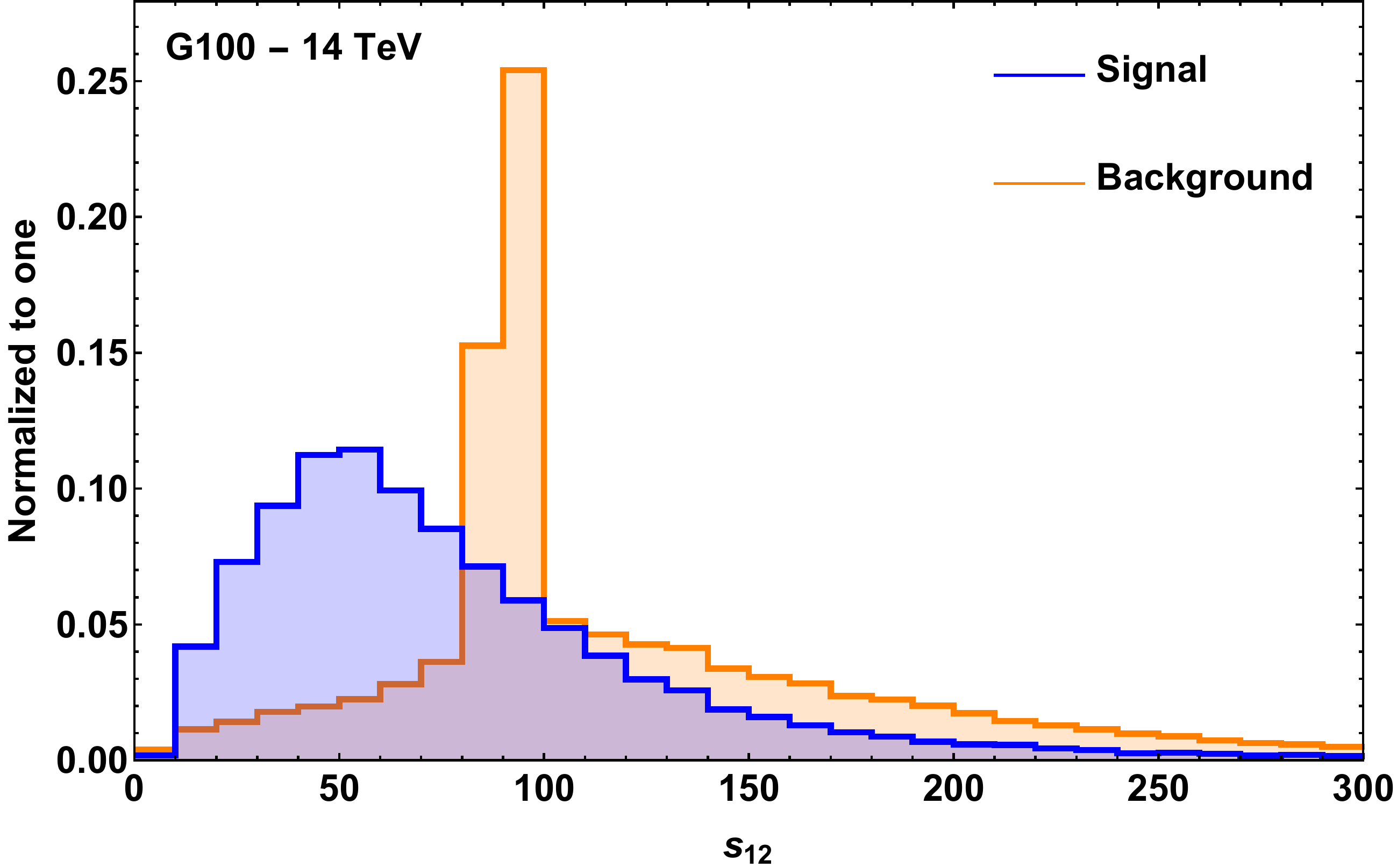}
	\includegraphics[width=7cm]{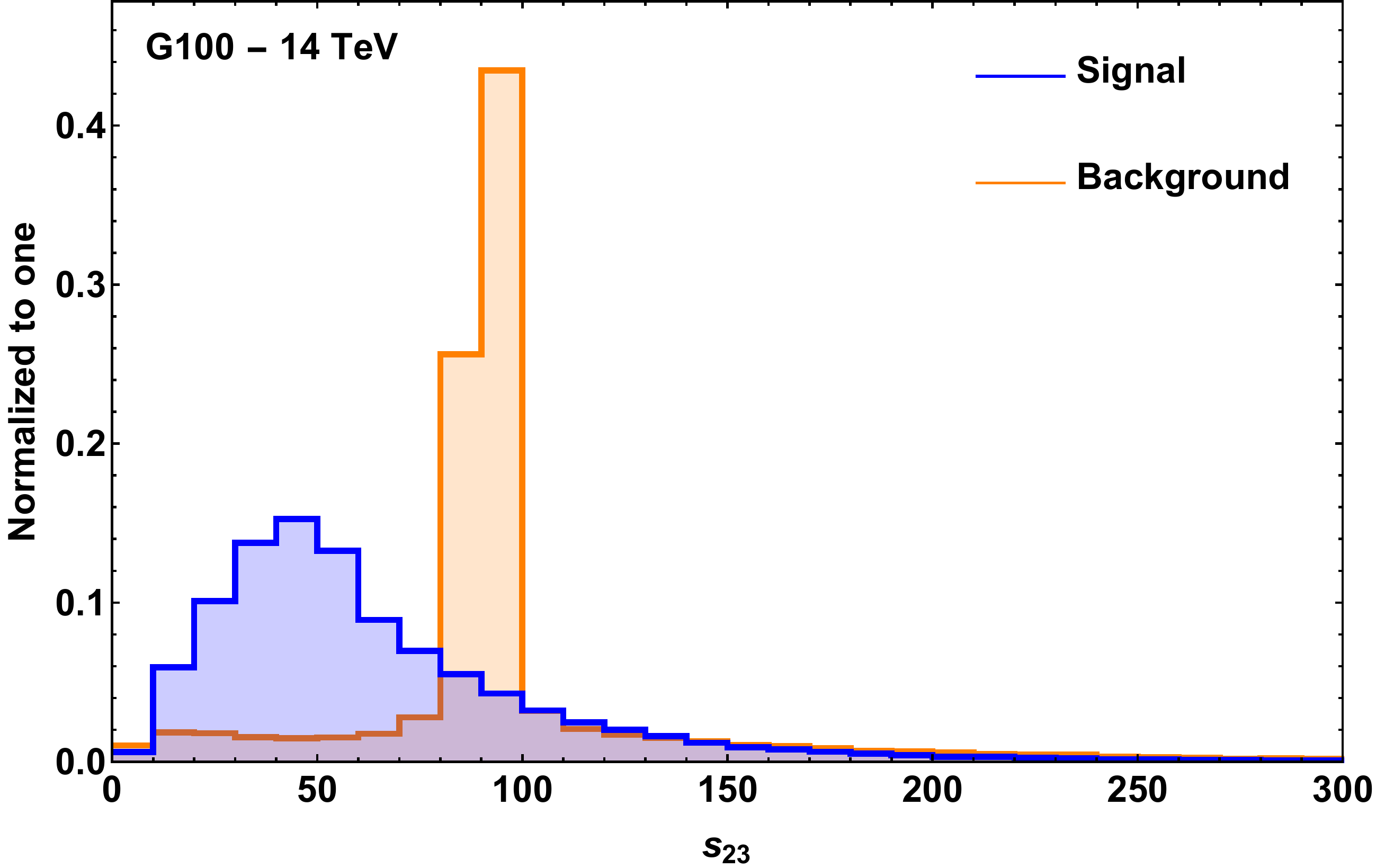}
	\caption{Distributions of CW neutrino production signal (blue) and background (orange) events in $s_{12}$ (left panel) and $s_{23}$ (right panel) at the LHC. The signal was simulated at the G100 benchmark point.}
	\label{fig:LHCplots}
\end{figure}

\begin{table}[t]
\begin{center}
	\begin{tabular}{|c||c|c||c|c|}
		\hline
		LHC - 14 TeV ($\mathcal{L} = 3000\ fb^{-1}$)  & \multicolumn{2}{c||}{\textbf{U100}} & \multicolumn{2}{c|}{\textbf{G100}} \\ \hline
		$\sigma$ (fb) with parton-level cuts        &   \multicolumn{2}{c||}{0.66}   &  \multicolumn{2}{c|}{1.39}     \\
		\# of signal events                         &   \multicolumn{2}{c||}{1965}   &  \multicolumn{2}{c|}{4180}     \\ \hline\hline
		Cuts:                                       &  S (fb) & BG (fb) &  S (fb) & BG (fb) \\ \hline 
		Pre-selection cuts                          &    0.29     &    93        &    0.62     &    93        \\
		$s_{12}<80$~GeV                             &    0.13     &    13.5        &    0.38     &    13.5        \\
		$s_{23}< 80$~GeV                            &    0.12     &    8.7         &    0.37     &    8.7         \\
		$M_{rec} < 120$~GeV                         &      --      &     --          &    0.29     &    3.8         \\ \hline\hline
		$S/B$                                       &   \multicolumn{2}{c||}{0.01}   &  \multicolumn{2}{c|}{0.077}    \\
		$S/\sqrt{B}$                                &   \multicolumn{2}{c||}{2.3}   &  \multicolumn{2}{c|}{8.2}     \\
		$S/\sqrt{S+B}$                              &   \multicolumn{2}{c||}{2.3}   &  \multicolumn{2}{c|}{7.9}     \\ \hline
	\end{tabular}
	\caption{Cut flow for the search for CW neutrino production in the $3\ell+\met$ channel at the HL-LHC (${\cal L}=3$ ab$^{-1}$). }
\label{tab:LHCcutflow}
\end{center}
\end{table}

\begin{figure}[t]
	\center
	\includegraphics[width=7cm]{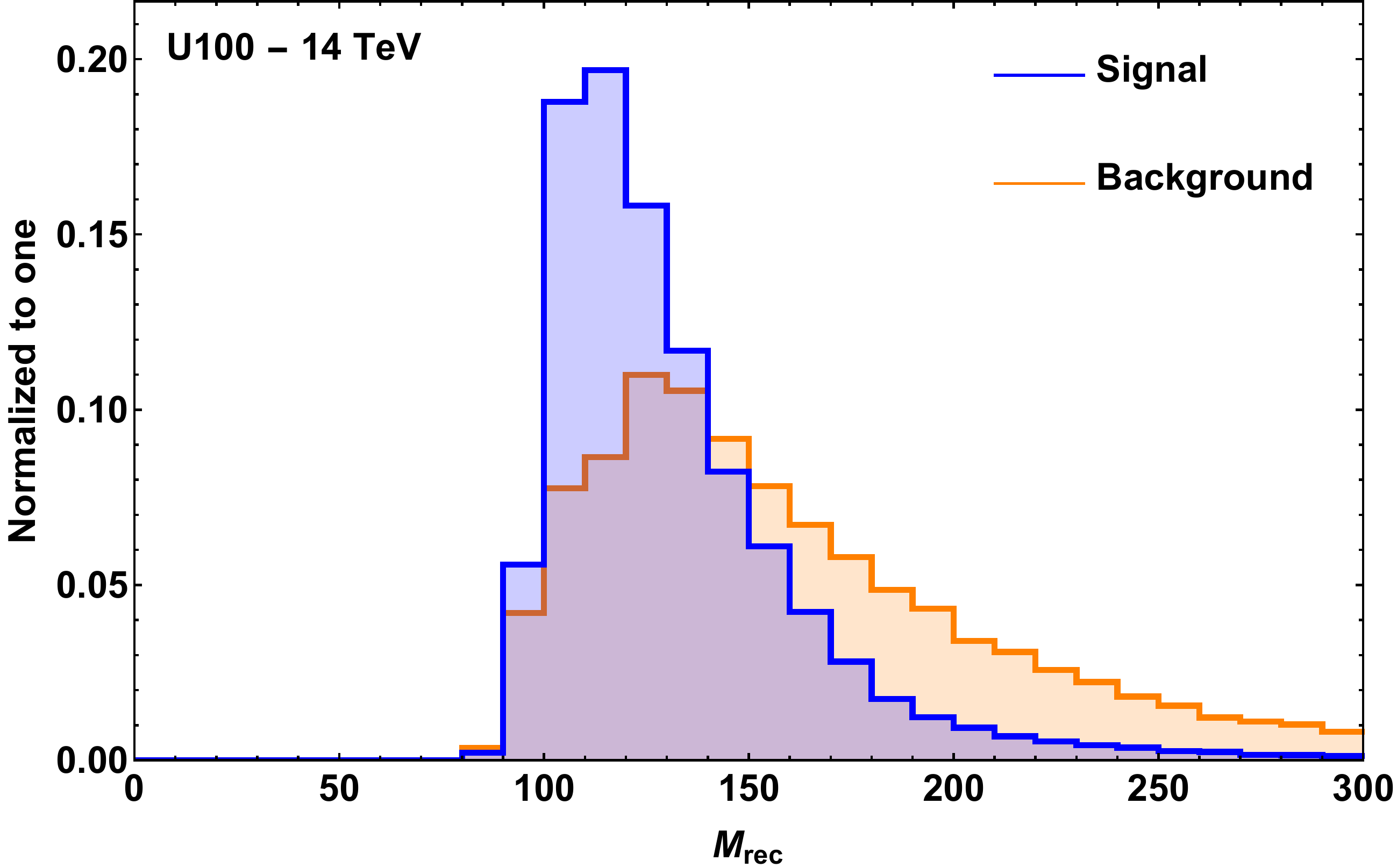}
	\includegraphics[width=7cm]{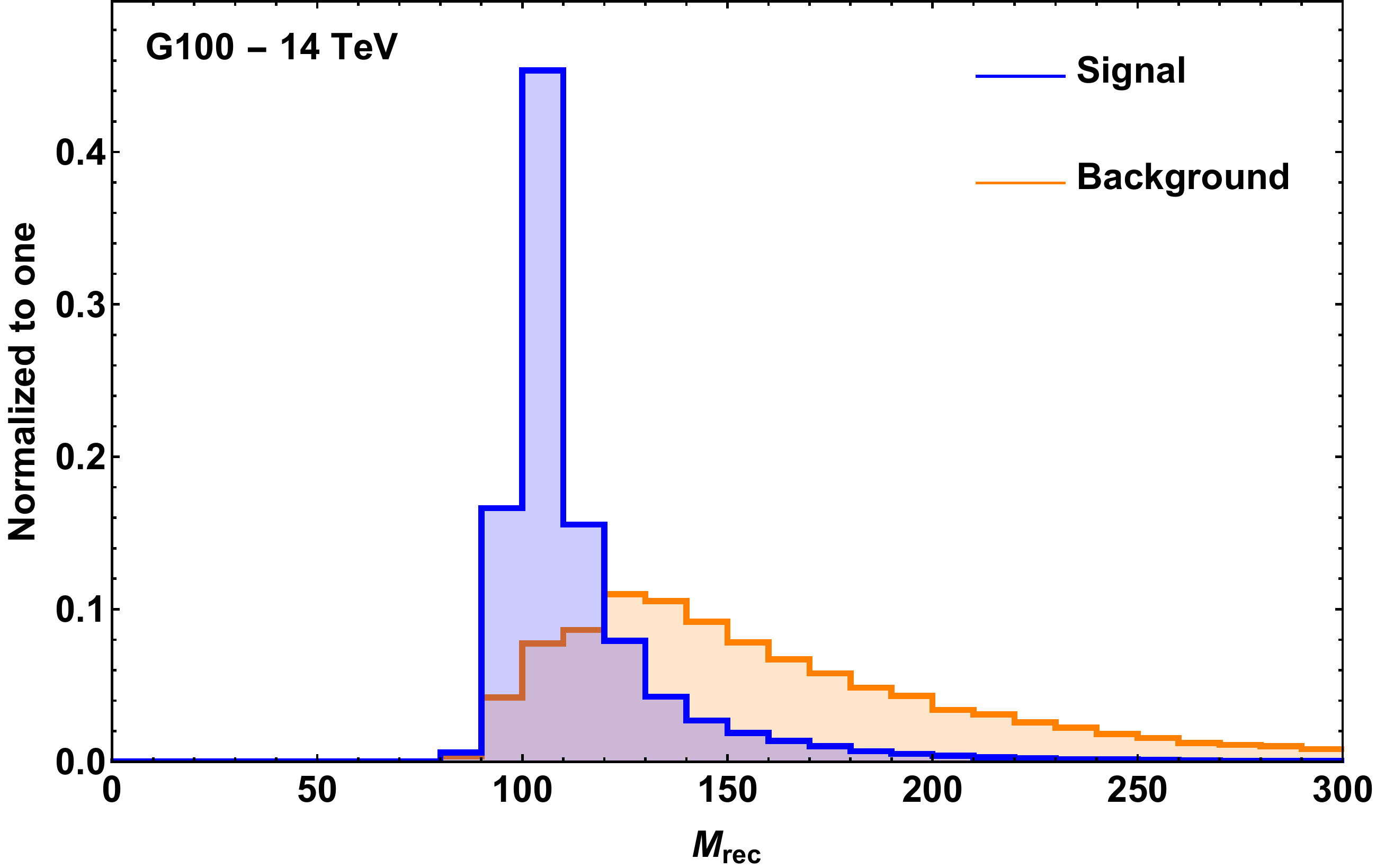}
	\caption{Distributions of CW neutrino production signal (blue) and background (orange) events in $M_{rec}$ at the U100 (left panel) and G100 (right panel) benchmark points.}
	\label{fig:LHCpeaks}
\end{figure}

Looking into the future, HL-LHC is expected to collect a ${\cal L}=3$ ab$^{-1}$ data set at $\sqrt{s}=14$~TeV. We have estimated the sensitivity of a simple search in the $3\ell + \met$ channel with this data set. For this search, events are processed as follows. First, if there is only one SFOC lepton pair in the event, the leptons in this pair are labeled 1 and 2, while the remaining (opposite-flavor) lepton is labeled 3. If there are multiple SFOC lepton pairs, the pair with the highest average $|p_T|$ is identified with leptons 1 and 2, and the remaining lepton is labeled 3. Furthermore, labels 1 and 2 are assigned so that the leptons 2 and 3 form an opposite-sign pair. With this labeling, lepton 1 predominantly corresponds to the particle produced in association woth the $\N$, lepton 2 to the particle produced directly in the decay $\N\to \ell W$, and lepton 3 to the particle produced in $W$ decay. Invariant masses of the opposite-sign lepton pairs, $s_{12}=(p_{\ell 1}+p_{\ell 2})^2$ and $s_{23}=(p_{\ell 2}+p_{\ell 3})^2$, are a useful signal discriminant, peaking sharply around $m_Z$ in the background (see Fig.~\ref{fig:LHCplots} and Table~\ref{tab:LHCcutflow}). Further, neutrino four-momentum $p_\nu$ can be fully reconstructed\footnote{A two-fold degeneracy is encountered when the quadratic $W$ mass constraint is used to determine the $z$ component of $p_\nu$. We follow the algorithm from Ref.~\cite{Chatrchyan:2011vp} to resolve this ambiguity. If there are two real solutions, we choose the solution with the smaller absolute value of $p_z$; if the solutions are complex, we use the real part as the $p_z$.} using the conservation of transverse momentum and requirements $p_\nu^2=0$ and $(p_\nu + p_{\ell 3})^2=m_W^2$. It can then be used to calculate the mass of the CW neutrino candidate, $M_{\rm rec}^2=(p_\nu + p_{\ell 2}+p_{\ell 3})^2$. In CW models with well-separated resonances, the signal appears as a sharp peak in this variable centered at the mass of the produced mode (or a series of peaks, if a number of modes can be produced with sizable rates). On the other hand, in CW models with closely-spaced resonances, the individual peaks are merged to a broader excess due to experimental resolution. This can be seen in Fig.~\ref{fig:LHCpeaks}, which compares the distributions in $M_{\rm rec}$ for uniform (U100) and generalized (G100) benchmark points. The sharp nature of the peak in the latter model allows for further background suppression using a cut on $M_{\rm rec}$. As a result, G100 model is easily discoverable at HL-LHC, while U100 may require a more refined analysis to be tested; see Table~\ref{tab:LHCcutflow}. Benchmark points with heavier $\N$ remain inaccessible at the LHC, even with the full HL-LHC data set, due to small production cross sections.      

\begin{table}[t]
	\begin{center}
		\begin{tabular}{ |c||r|r|r|r|r|r|r| }
			\hline
			BP &  U100 & G100 & U400 & G300 & U750 & G750 & U1000  \\ \hline
			$\sqrt{s}$, GeV & 250 & 250 & 500 & 500 & 3000 & 3000 & 3000  \\
			${\cal L}_{3\sigma}$, fb$^{-1}$ & 220 & 50 & 4300 & 20 & 55 & 25 & 720 \\ \hline
		\end{tabular}
		\caption{Center-of-mass energy and integrated luminosity required for a 3-sigma observation of the CW neutrino signal in electron-positron collisions.}
		\label{tab:LEPTONlums}
	\end{center}
\end{table}

We have also studied the prospects for proposed future lepton colliders, such as the ILC~\cite{Baer:2013cma,Fujii:2017vwa}, CEPC~\cite{CEPCStudyGroup:2018ghi}, FCC-ee~\cite{Mangano:2018mur,Benedikt:2018qee} and CLIC~\cite{Aicheler:2012bya,Charles:2018vfv}. We analyzed three center-of-mass energies: $\sqrt{s}=250$~GeV, 500~GeV, and 3~TeV. Signal cross sections for the 7 benchmark points, listed in Table~\ref{tab:Xsections}, are in the $1-10$ fb range in all cases where CW neutrinos are kinematically accessible. This implies that ${\cal O}(10^3-10^4)$ signal events would be collected with data sets envisioned at these colliders. Since hadronic backgrounds are much less of an issue at lepton colliders, we focus on hadronic $W$ decays, {\it i.e.} the final state $\ell \nu j j$. SM backgrounds were simulated inclusively; the main irreducible background is $e^+e^-\to W^+W^-$, with one leptonic and one hadronic $W$ decay. We find that in all cases, a simple cut-based analysis is sufficient to observe the signal with realistic luminosities; see Table~\ref{tab:LEPTONlums}. Details of the analysis are discussed in Appendix~\ref{sec:LCapp}.

\begin{figure}[t!]
	\centering
	\includegraphics[width = 5 cm]{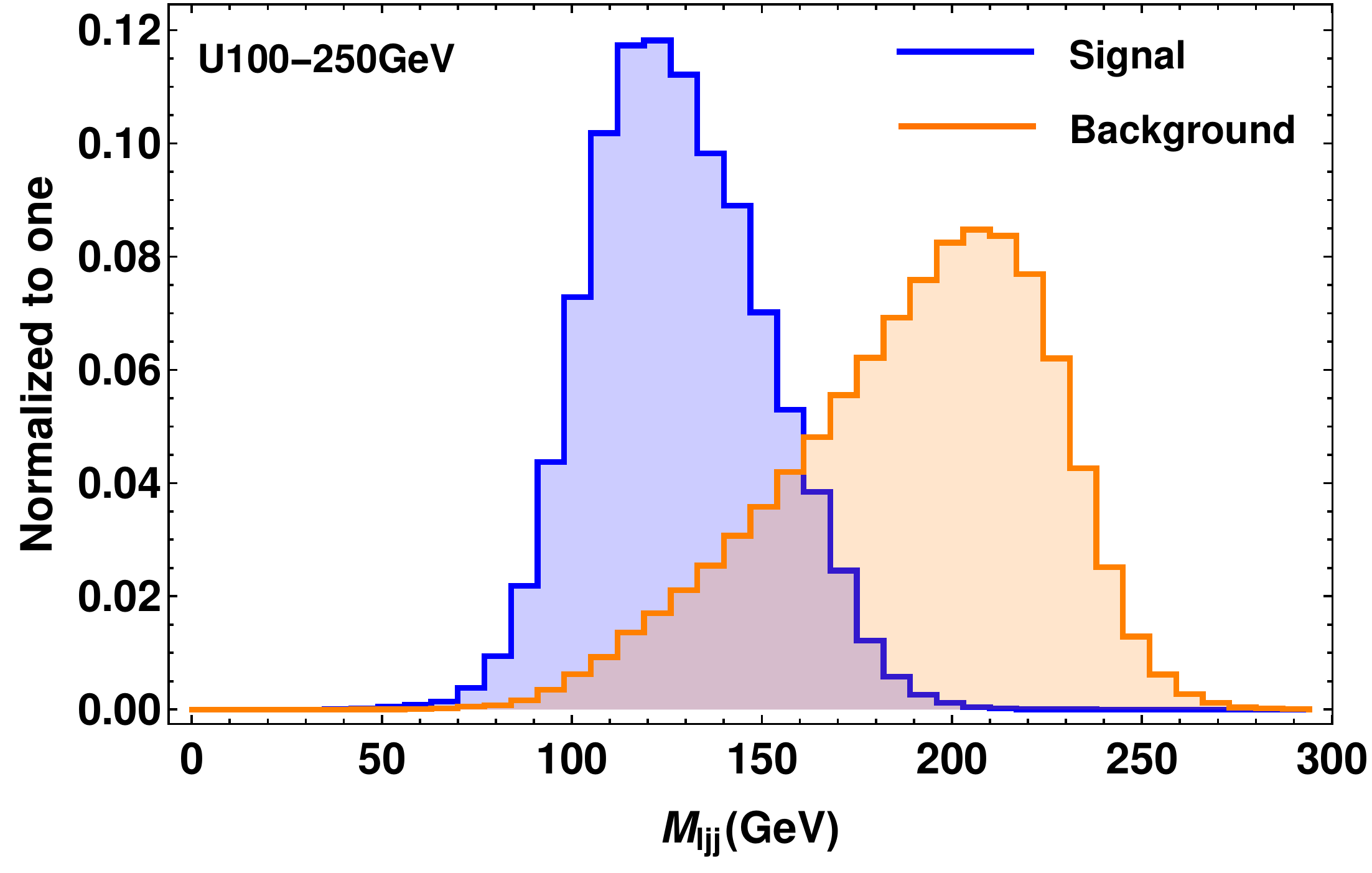}
	\includegraphics[width = 5 cm]{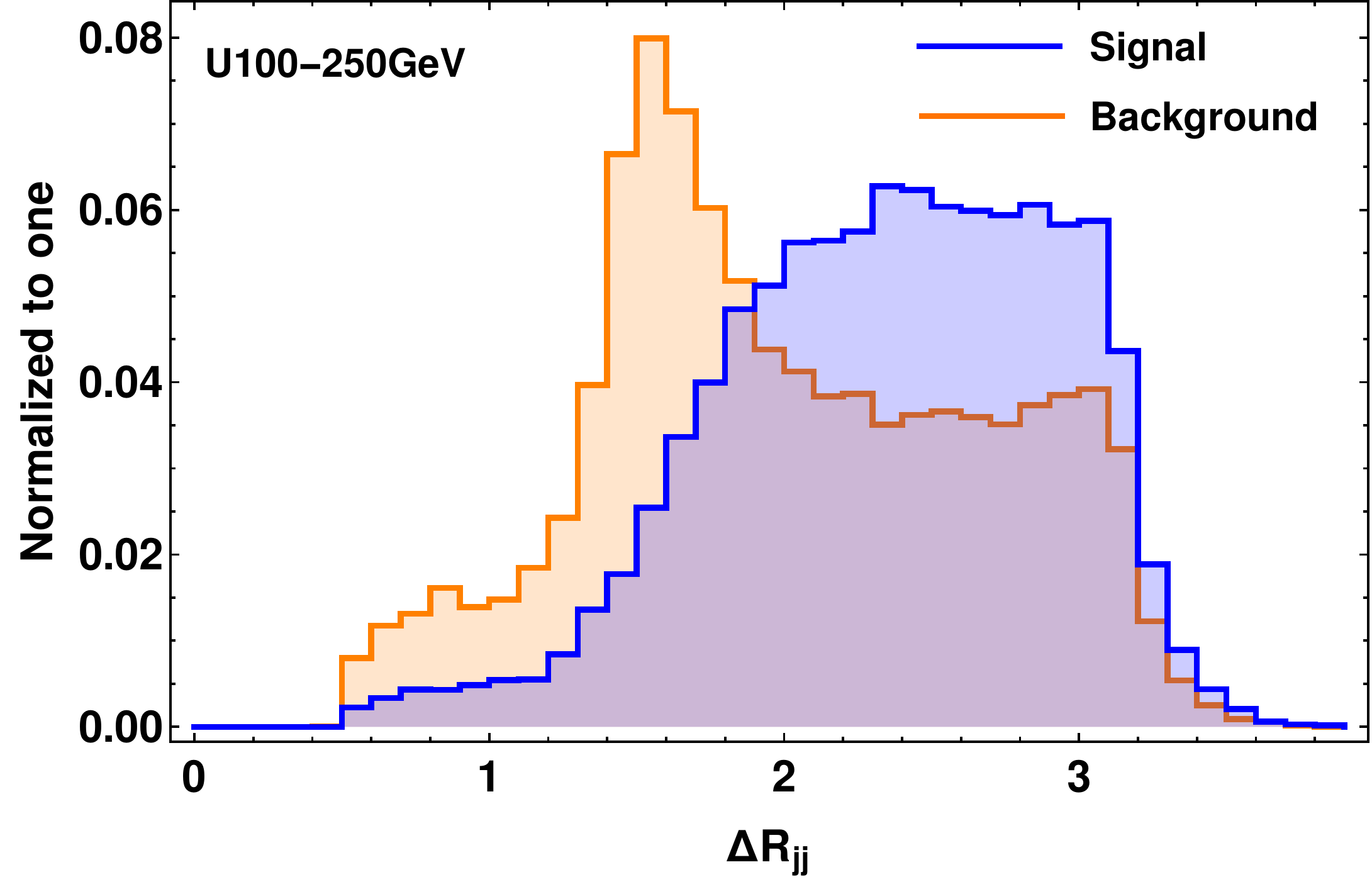}
	\includegraphics[width = 5 cm]{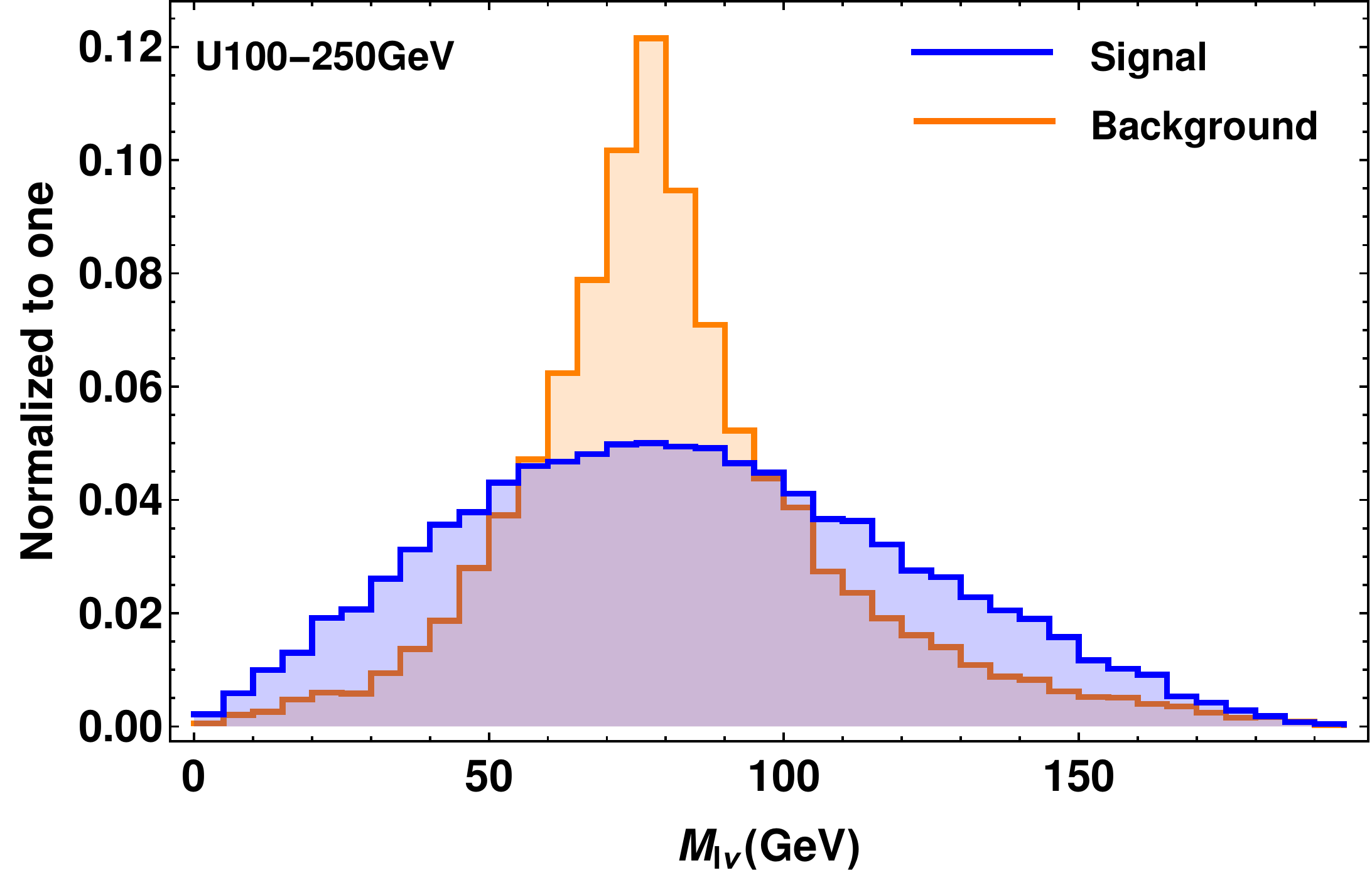}
	\caption{Signal (blue) and background (orange) distributions for U100 model at a 250 GeV $e^+e^-$ collider: $M_{\ell j j}$ (left) after pre-selection cuts, $\Delta R_{jj}$ (middle) after $M_{\ell j j}$ cut, and $M_{\ell \nu}$ (right) after $\Delta R_{jj}$ cut.   
	}
	\label{fig:ILCdists}
\end{figure}%

\begin{table}[t]
	\centering
	\begin{tabular}{|c|c|c|}
		\hline 
		\multicolumn{3}{|c|}{\bf{U100}} \\
		\hline
		\multicolumn{3}{|c|}{$\sqrt{s}=250$~GeV, $\mathcal{L}=2000$ fb$^{-1}$} \\
		\hline
		Cuts & S (fb) & BG (fb)  \\
		\hline \hline
		Parton-level cuts & 7.4 & 3200  \\
		\hline
		Pre-selection cuts & 3.2 & 1650 \\
		\hline
		$ M_{\ell j j} \in [70, 140]$~GeV & 2.3 & 160  \\
		$ \Delta R_{jj} \in [1.8, 3.5]$ & 1.9 & 91\\
		$ M_{\ell \nu} \cancel{\in} [60, 90]$~GeV & 1.3 & 43  \\
		\hline
		$S/B$ & \multicolumn{2}{|c|}{0.03} \\
		$S/\sqrt{B}$ & \multicolumn{2}{|c|}{9.0}  \\
		$S/\sqrt{S+B}$ & \multicolumn{2}{|c|}{8.9} \\
		\hline
	\end{tabular}  \\
	\vspace{0.5cm}
	\caption{Cut flow table for the search for CW neutrinos at the U100 benchmark point, in $e^+e^-$ collisions at $\sqrt{s}=250$~GeV.}
	\label{tab:ILCcuts}
\end{table} 

As an example, consider the case of U100 benchmark point, which as discussed above is non-trivial to observe at the LHC even with the full HL-LHC data set. At electron-positron colliders, $\sqrt{s}=250$ GeV is sufficient to produce CW neutrinos at the U100 benchmark point.  Fig.~\ref{fig:ILCdists} shows distributions of signal and background events in three variables that are useful for signal/background discrimination, $M_{\ell jj}=\sqrt{(p_\ell + p_{j1}+p_{j2})^2}$, $\Delta R_{jj}$, and $M_{\ell \nu}=\sqrt{(p_\ell+p_\nu)^2}$. Here $p_\nu$ is the neutrino 4-momentum, reconstructed from the three-dimensional missing momentum supplemented with the condition $m_\nu=0$. The signal distribution in $M_{\ell jj}$ results from a number of peaks, corresponding to different CW modes, merged into a continuous ``hump" due to experimental smearing effects. The larger values of $\Delta R_{jj}$ for the signal events are due to the fact that for a 100 GeV CW neutrino, the $W$ boson is almost exactly at rest in the lab frame. Finally, the distributions in $M_{\ell \nu}$ reflect that fact that almost all $\ell\nu$ pairs in the background come from a single $W$ decay, while in the signal this is not the case. A series of cuts in these three variables, summarized in Table~\ref{tab:ILCcuts}, is sufficient for a 3-sigma observation of the signal with about 220 fb$^{-1}$, ignoring systematic errors. The required integrated luminosity is far below the $2-5$ ab$^{-1}$ projected at the proposed $e^+e^-$ Higgs factories. With the full projected data sets, such colliders can perform detailed measurements to uncover the nature of the signal. For example, the shape of the signal distribution in $M_{\ell jj}$ can be used to distinguish the CW neutrino tower from a single massive sterile neutrino state appearing in other models. We will study the details of this measurement in future work.           

\section{Conclusions and Outlook}
\label{sec:conc}

In this paper, we investigated complete, fully realistic models which produce small Dirac neutrino masses without unnaturally small parameters using the clockwork mechanism. The main results can be summarized as follows:

\begin{itemize}
	\item In the uniform clockwork model, a perturbation theory was developed and applied to obtain approximate analytic expressions for quantities of phenomenological interest, such as excited CW neutrino masses and couplings;
	\item Experimental constraints on the uniform model from flavor-changing decay $\mu\to e\gamma$ and precision electroweak fits were calculated. It was found that CW neutrinos can have masses in the 100 GeV -- 1 TeV range, within reach of the LHC and proposed lepton colliders, with neutrino Yukawa couplings of order $10^{-1}-10^{-2}$;
	\item It was shown that the uniform clockwork model is only one representative of a much more general class of models that implement the clockwork mechanism for neutrino masses. Phenomenology of two sample generalized CW models was studied;
	\item Collider signatures of CW neutrinos in the uniform and a generalized CW models were studied using Monte Carlo simulations of signal and background. It was found that at the LHC, models with light ($\sim 100$~GeV) CW neutrinos can be discovered using the $3\ell+\met$ signature, although the integrated luminosities required are larger than what has been collected so far. Lepton colliders will be able to discover the CW neutrinos as long as they are within their kinematic range.     
\end{itemize}

In the future, we would like to extend the analysis of this paper in several directions. First, as already mentioned, it would be interesting to understand whether and how collider experiments can distinguish the CW model from a more traditional model with a single heavy neutrino. Second, throughout this paper we chose the CW mass scale $m$ to be in the neighborhood of the weak scale. This is motivated by simplicity and by our interest in collider phenomenology, but is by no means required by the models themselves. (Note that the light neutrino mass, Eq.~\leqn{zeromode_mass}, is to leading order completely independent of $m$.) It would be interesting to investigate phenomenologically a broader range of the CW scales, which may entail constraints and signatures different from the ones considered here.

\section*{Acknowledgements}

We are grateful to Yuval Grossman, Aaron Pierce and Yue Zhao for useful discussions related to this work. This research is supported by the U.S. National Science Foundation through grant PHY-1719877, and by Cornell University through the Bethe Postdoctoral Fellowship (SH). 

\appendix
\section{Perturbation Theory in $p$}
\label{app:PT}

In this Appendix, we will compute eigenvalues and eigenvectors to the clockwork mass matrix after including the Yukawa coupling to the Standard Model sector at the zeroth site, for the uniform clockwork model of Section~\ref{sec:base}. They are already known exactly for the $y = 0$ case where the clockwork modes are completely decoupled from the SM. For the $y > 0$ case, we can compute them to leading order using perturbation theory in $p = \frac{y v}{\sqrt{2} m}$ as shown below. In the following, matrices and vectors with tildes are unperturbed (with $y = 0$) and those without tildes are perturbed. For the eigenvalues, we use $\bar{\lambda}_i$ to refer to the perturbed eigenvalues, and $\lambda_i$ are quantities defined in Eq.~(\ref{masses}) which are the eigenvalues of the unperturbed squared matrix.

\subsection{Eigenvalues}

If we consider $M^\dag M$ and its eigenvectors, the unperturbed (right) eigenvectors are given by $\tilde{w}^{(i)}_j = \tilde{U}_R^{ji}$, where the superscript $(i)$ denotes the mode number, while the subscript $j$ is the vector component. On including the Yukawa coupling, the perturbation to this matrix has zeroes in all entries except for $(\delta M^2)^{00} = p^2$:
\begin{align}
\frac{M^\dag M}{m^2} = 
\begin{pmatrix}
p^2 + q^2 & -q  & \cdots & 0 & 0 \\
-q & 1 + q^2 & \cdots & 0 & 0 \\
\vdots & \vdots & \ddots & \vdots & \vdots \\
0 & 0 & \cdots & -q & 1
\end{pmatrix}
= \underbrace{\begin{pmatrix}
	q^2 & -q  & \cdots & 0 & 0 \\
	-q & 1 + q^2 & \cdots & 0 & 0 \\
	\vdots & \vdots & \ddots & \vdots & \vdots \\
	0 & 0 & \cdots & -q & 1
	\end{pmatrix}}_{\mathrm{unperturbed\ } \frac{\tilde{M}^\dag \tilde{M}}{m^2}}
+ \underbrace{\vphantom{\begin{pmatrix} \ddots \\ \ddots \\ \ddots \\ \end{pmatrix}} 
	\begin{pmatrix}
	p^2 & & \mathbf{0}_{1\times N}  \\
	& &                         \\    
	\mathbf{0}_{N\times 1}   & & \mathbf{0}_{N\times N}  \\ 
	\end{pmatrix}}_{\frac{\delta M^2}{m^2}}
\end{align}
Thus, to first order in perturbation theory in $p$, the shift in eigenvalues can be written as follows (note that $\lambda_i$ are the eigenvalues of the squared matrix $\tilde{M}^\dag \tilde{M}$ and the actual masses of the neutrinos will be $m_i = m \sqrt{\lambda_i}$): 
\beq 
\delta\lambda_i = \left\langle w^{(i)} \right| \frac{\delta M^2}{m^2} \left| w^{(i)} \right\rangle = p^2 (\tilde{U}_R^{0i})^2
\eeq{pt1}
This gives the results in Eq.~(\ref{zeromode_mass}) and (\ref{ith_dude_mass}). The only difference is the $\mathcal{O}(p^3)$ term in Eq.~(\ref{zeromode_mass}). This term requires a higher order calculation using the determinant form of the eigenvalue equation, which we do not expand on here. 

These results can be cross-checked by considering the unperturbed left-handed eigenvectors as well, but the calculation in this case is not as straightforward, because the perturbation to $\tilde{M} \tilde{M}^\dag$ has more matrix entries, with some at $\mathcal{O}(p)$. The leading order ($\mathcal{O}(p)$) perturbation calculation in the left-handed case results in zero deviation in eigenvalues, since the first non-trivial deviations appear at $\mathcal{O}(p^2)$. This necessitates a second-order calculation. We have performed this calculation and confirmed explicitly that $M M^\dag$ and $M^\dag M$ have identical eigenvalues.

\subsection{Eigenvectors and Rotation Matrices}

Once the eigenvalues are known, it is relatively straightforward to calculate the perturbed eigenvectors, and thus the rotation matrices that diagonalize the perturbed mass matrix. For both left and right eigenvectors, it is more convenient to do this in the basis in which the unperturbed clockwork mass matrix is diagonal. Thus, we consider the eigenvalue equations for the matrices $\tilde{U}_L^\dag (M M^\dag) \tilde{U}_L$ for the left-handed case, and $\tilde{U}_R^\dag (M^\dag M) \tilde{U}_R$ for the right-handed case. We will refer to right eigenvectors as $w^{(i)}$ and left eigenvectors as $v^{(i)}$ in this basis. The rotation matrices in our convention have the eigenvectors as its column vectors: 
\bea
U_L^{ij} = (\; \tilde{U}_L(\mathbf{I}+\Delta_L) \;)^{ij} = (v_o^{(j)})_i \\
U_R^{ij} = (\; \tilde{U}_R(\mathbf{I}+\Delta_R) \;)^{ij} = (w_o^{(j)})_i
\eea
where $\mathbf{I}$ is a unit matrix, and $v_o$ and $w_o$ are the eigenvectors in the \textit{original} clockwork basis.

In the \textquoteleft clockwork-diagonal' basis (where the $p = 0$ mass matrix is diagonal), the unperturbed eigenvectors are columns of the identity matrix, i.e. $\tilde{v}^{(i)}_j = \tilde{w}^{(i)}_j = \delta_{ij}$. Thus, we write the perturbed eigenvectors as,
\bea
(\mathbf{I}+\Delta_L)^{ij} = v^{(j)}_i \\
(\mathbf{I}+\Delta_R)^{ij} = w^{(j)}_i
\eea
We will solve for $v^{(i)}$ and $w^{(i)}$ in the rest of this section. $\Delta_L$ and $\Delta_R$ are related to $v^{(i)}$ and $w^{(i)}$ as mentioned above, and these are the quantities shown in Section~\ref{sec:base}, in Eqs.~(\ref{Rs}) and (\ref{Ls}).

\textbf{Left eigenvectors:}

In the basis in which the unperturbed clockwork mass matrix is diagonal, the eigenvalue equation for $MM^\dag$ can be written as follows:
\beq
\left( \begin{pmatrix}
	0           &           &        &                       \\
	& \lambda_1 &        &                       \\
	&           & \ddots &                       \\
	&           &        & \lambda_N 
\end{pmatrix}
+
\begin{pmatrix}
	p^2 & & W^\dag                  \\
	& &                         \\    
	W   & & \mathbf{0}_{N\times N}  \\ 
\end{pmatrix}
- \bar{\lambda}_i \mathbf{I}_{N+1} \right) v^{(i)} = 0
\eeq{whatever}
where $\lambda_i$ is the unperturbed eigenvalue and $\bar{\lambda}_i$ is the exact eigenvalue. $W$ is a column vector obtained from rotating the perturbation terms in the matrix to this basis. Its terms are given by
$$ W_j = -\sqrt{\frac{2 p^2 q^2}{N+1}} \sin \frac{N j \pi}{N+1} = - \sqrt{\frac{p^2}{N+1} \lambda_j C_j}$$
Decomposing the eigenvalues and eigenvectors into sums of unperturbed and perturbed parts, we have,
\begin{align}
\bar{\lambda}_i &= \lambda_i + \Delta_i \\
v^{(i)}_j &= \tilde{v}^{(i)}_j + \delta v^{(i)}_j = \delta_{ij} + \delta v^{(i)}_j
\end{align}
where from the previous section, we know the values of $\Delta_i = \delta \lambda_i$.

The unperturbed parts in the eigenvalue equation give a zero on multiplying $\delta_{ij}$. Thus, the equation becomes,
\beq
\begin{pmatrix}
	0-\lambda_i &                       &        &                       \\
	& \lambda_1 - \lambda_i &        &                       \\
	&                       & \ddots &                       \\
	&                       &        & \lambda_N - \lambda_i 
\end{pmatrix} \delta v^{(i)}
+
\left( \begin{pmatrix}
	p^2 & & W^\dag                  \\
	& &                         \\    
	W   & & \mathbf{0}_{N\times N}  \\
\end{pmatrix}
- \delta \lambda_i \mathbf{I}_{N+1} \right) (\tilde{v}^{(i)}+\delta v^{(i)}) = 0
\eeq{app1}

This simplifies to the following equations:
\bea
\mathbf{i=0,\ j=0:}          &\ \  (p^2 - \Delta_0)(1 + \delta v^{(0)}_0) + \sum_{k=1}^N W_k \delta v^{(0)}_k = 0 \label{eq:LH_eigenvector_1}  \\
\mathbf{i=0,\ j\neq 0:}      &\ \  \lambda_j \delta v^{(0)}_j + W_j(1 + \delta v^{(0)}_0) - \Delta_0 \delta v^{(0)}_j = 0 \label{eq:LH_eigenvector_2} \\
\mathbf{i>0,\ j=0:}          &\ \  -\lambda_i \delta v^{(i)}_0 + W_i + (p^2 - \Delta_i)\delta v^{(i)}_0 + \sum_{k=1}^N W_k \delta v^{(i)}_k = 0 \label{eq:LH_eigenvector_3} \\
\mathbf{i>0,\ j=i:}          &\ \  W_i \delta v^{(i)}_0 - \Delta_i (1 + \delta v^{(i)}_i) = 0 \label{eq:LH_eigenvector_4} \\
\mathbf{i, j>0, j\neq i:} &\ \  (\lambda_j-\lambda_i) \delta v^{(i)}_j + W_j \delta v^{(i)}_0 - \Delta_i \delta v^{(i)}_j = 0 \label{eq:LH_eigenvector_5}
\eea

On solving these equations, dropping higher order terms and plugging in the appropriate $\delta \lambda_i$ when necessary, we get the following results:
\begin{align}
v^{(0)}_j &= -\frac{W_j}{\lambda_j}\ ,\ j>0 \ \ \ (\text{from Eq.~(\ref{eq:LH_eigenvector_2})}) \\
v^{(i)}_0 &= - v^{(0)}_i = \frac{W_i}{\lambda_i}\ ,\ i>0 \ \ \ (\text{from Eq.~(\ref{eq:LH_eigenvector_4})}) \\
v^{(i)}_j &= \frac{W_i W_j}{\lambda_i(\lambda_i-\lambda_j)}\ ,\ j>0,\ j\neq i \ \ \ (\text{from Eq.~(\ref{eq:LH_eigenvector_5})}) 
\end{align}
These results are consistent with equations Eq.~(\ref{eq:LH_eigenvector_1}) and Eq.~(\ref{eq:LH_eigenvector_3}) up to $\mathcal{O}(p^2)$.\footnote{In showing the consistency with Eq.~(\ref{eq:LH_eigenvector_1}), one may need the following identity, which can be proven using Eqs.~(1.447.3) and~(1.353.3) of Ref.~\cite{GR}. 
	\bea
	\sum_{j=1}^N \frac{\sin^2 \frac{j N \pi}{N+1}}{\lambda_j} &=& \frac{N+1}{2 q^2} \left( 1 - \frac{1}{q^{2N}} \left( \frac{q^2-1}{q^2-q^{-2N}} \right) \right).
	\eea
}

The equations above do not constrain the deviations in the diagonal elements of the left-handed vectors, $\delta v^{(i)}_i$, as all terms containing them enter only at higher order. These can be obtained from the normalization of the vectors, or in other words, from the unitarity of the left-handed rotation matrix. To do so, it is important to note that the deviations in (left) eigenvectors are not of the same order. The pseudozero mode eigenvector has deviations of $\mathcal{O}(p)$, while heavier mode eigenvectors have deviations of $\mathcal{O}(p^2)$. To start with, the unitarity of $U_L$ implies
\bea
\Delta_L^\dagger + \Delta_L + \Delta_L^\dagger \Delta_L = 0
\eea
Expanding $\Delta_L = p \mathcal{O}_1 + p^2 \mathcal{O}_2 + \mathcal{O} (p^3)$ then implies that 
\bea
&& p(\mathcal{O}_1^\dag+\mathcal{O}_1)+p^2(\mathcal{O}_2^\dag + \mathcal{O}_2 + \mathcal{O}_1^\dag\mathcal{O}_1) + \cdots = 0 \\
\Rightarrow && \mathcal{O}_1^\dag + \mathcal{O}_1 = 0  \ \ ; \ \ \mathcal{O}_2^\dag + \mathcal{O}_2 + \mathcal{O}_1^\dag\mathcal{O}_1 = 0.
\eea
In terms of eigenvector deviations, this implies,
\begin{align}
&\delta v^{(0)}_i + \delta v^{(i)}_0 = 0, \text{ for } i \neq 0 \\
\delta v^{(i)}_j +& \delta v^{(j)}_i + \delta v^{(0)}_i \delta v^{(0)}_j = 0, \text{ for } i,j \neq 0 \\
\delta v^{(0)}_0 +& \delta v^{(0)}_0 + \sum_{i=1}^N (\delta v^{(0)}_i)^2 = 0
\end{align}
The correction to $v^{(i)}_i$ and $v^{(0)}_0$ can thus be determined using the unitarity of the rotation matrix, and we get, 
\begin{align}
& \delta v^{(i)}_i = -\frac{1}{2}\left( \delta v^{(0)}_i\right)^2, \text{ for } i \neq 0  \\
& \delta v^{(0)}_0 = -\frac{1}{2}\sum_{i=1}^N (\delta v^{(0)}_i)^2
\end{align}
In summary, to leading order, the left eigenvectors are given by,
\begin{align}
v^{(0)}_0 &= 1 - \frac{p^2}{N+1} \sum_{i=1}^N \frac{C_i}{2\lambda_i}\\
v^{(0)}_i &= -v^{(i)}_0 = \sqrt{\frac{p^2}{N+1}\frac{C_i}{\lambda_i}} \\
v^{(i)}_i &= 1 - \frac{p^2}{N+1} \frac{C_i}{2\lambda_i}\\
v^{(i)}_j &= \frac{p^2}{N+1} \sqrt{\frac{\lambda_j}{\lambda_i}} \frac{\sqrt{C_i C_j}}{\lambda_i-\lambda_j}
\end{align}

\textbf{Right eigenvectors:}

As seen before, the perturbation matrix in the right-handed case (i.e. $M^\dag M$) is simpler than in the left-handed case. In the \textquoteleft clockwork-diagonal' basis, we get the following eigenvalue equation:
\begin{align*}
(D + zz^\dag-\bar{\lambda}_i \mathbf{I}_{N+1} ) w^{(i)}= 0
\end{align*}
where D is the diagonal matrix of unperturbed eigenvalues $= \mathrm{diag}(\lambda_0, \lambda_1 \cdots)$ and $z$ is a vector given by $z = \tilde{U}_R^\dag\; (p, 0, \cdots)^T$. In terms of previously defined quantities, the column vector $z$ simplifies to
$$
z=\left(p\ \sqrt{\frac{q^2-1}{q^{2(N+1)}-1}},\ \sqrt{\frac{p^2}{N+1}}\sqrt{C_i}\right)^T
$$
Using a perturbative expansion as in the case of the left eigenvectors, the eigenvalue equation becomes,
\begin{align*}
&(D + zz^\dag-(\lambda_i+\Delta_i) \mathbf{I}_{N+1} ) (\tilde{w}^{(i)}+\delta w^{(i)}) = 0 \\
\Rightarrow &\ (D - \lambda_i I_{N+1} ) \delta w^{(i)} + (zz^\dag - \Delta_i I_{N+1})(\tilde{w}^{(i)}+\delta w^{(i)}) = 0 
\end{align*}
The $j^{th}$ row of this equation gives,
\begin{align}
(\lambda_j - \lambda_i)\delta w^{(i)}_j + \sum_{k=0}^N z_j z_k (\delta_{ik}+\delta w^{(i)}_k)-\Delta_i(\delta_{ij}+\delta w^{(i)}_j)=0
\end{align}
where the substitution $\tilde{w}^{(i)}_j = \delta_{ij}$ has been made.

The corrections to the eigenvalues $\Delta_i$ are known from previous calculations and can be written in terms of $z$ as $\Delta_i =  z_i^2$. Solving for $\delta w^{(i)}_j$ for various cases as before, we get the following:
\begin{align}
& w^{(i)}_i = 1 \\
& w^{(0)}_j = - w^{(j)}_0 = -\frac{z_0 z_j}{\lambda_j} = - p^2 \sqrt{\frac{1}{N+1}\ \frac{q^2-1}{q^{2(N+1)}-1}} \frac{\sqrt{C_j}}{\lambda_j}\\
& w^{(i)}_j = \frac{z_i z_j}{\lambda_i - \lambda_j}  = \frac{p^2}{N+1} \frac{\sqrt{C_i C_j}}{\lambda_i - \lambda_j}.
\end{align}

\section{Lepton Collider Analysis Details} 
\label{sec:LCapp}

In this Appendix, we present a brief summary of lepton collider analysis for all seven benchmark points introduced in Section.~\ref{sec:coll}. Choices of theory parameters for each of the benchmarks are given in Table.~\ref{tab:UBPs} and Table.~\ref{tab:GBPs}. The relevant process at lepton colliders is shown as a Feynman diagram in Fig.~\ref{fig:collider_signal_process}. Considered center-of-mass energies and integrated luminosities are $\sqrt{s} = 250$ GeV ($\mathcal{L}=2000$ fb$^{-1}$), 500 GeV ($\mathcal{L}=4000$ fb$^{-1}$), and 3 TeV ($\mathcal{L}=2000$ fb$^{-1}$).

We imposed $p_{T,\ell} >20$ GeV and $p_{T,j} > 20$ GeV cuts for both signal and background events simulations. We then imposed the following cuts on the fully processed (i.e. after hadronization (via \texttt{Pythia8}) and detector effects (via \texttt{Delphes3})), data as our event selection cuts (called ``pre-selection cuts''):
\bea
&& N_{\ell} \geq 1, \; N_{j} \geq 2 \nonumber \\
&& | \eta_\ell | < 2.5, \; | \eta_j | < 2.5 \nonumber \\
&& \Delta R_{\ell j} > 0.4, \; \Delta R_{j j} > 0.4 \\
&& p_{T, \ell} > 20 (1+r) \; {\rm GeV}, \; p_{T, j} > 20 (1+r) \; {\rm GeV}, \nonumber 
\label{eq:presel_cuts}
\eea
where $N_{\ell (j)}$ denotes the number of leptons (jets) in the event and $0< r < 1$ parametrizes smearing by detector effects. The size of $r$ will primarily be determined by the jet $E$ resolution of the detector, and typically for lepton colliders it will be $\sim \mathcal{O} (5) \%$. We, however, will use $25 \%$ for most of our analysis to make a conservative estimation. The only exception will be for \textbf{G100} and \textbf{U100} where the majority of signal events are distributed near the low $p_T$ regime and hence we use $5 \%$ to secure enough signal events.

We summarize the list of kinematic cuts and their efficiency in Table~\ref{tab:cutflow_lepton_colliders_generalized_model_1} and Table~\ref{tab:cutflow_lepton_colliders_generalized_model_2} for generalized CW models and in Table~\ref{tab:cutflow_lepton_colliders_uniform_model_1} and Table~\ref{tab:cutflow_lepton_colliders_uniform_model_2} for uniform models. We denote the jet with highest $p_T$ as $j_1$.


%
%
\begin{table}[ht]
	\centering
	\begin{tabular}{|c|c|c|c|c|c|c|}
		\hline 
		\multicolumn{7}{|c|}{ \textbf{G100} } \\
		\hline
		-- & \multicolumn{2}{c|}{$\sqrt{s}= 250$ GeV} &  \multicolumn{2}{c|}{$\sqrt{s}= 500$ GeV} & \multicolumn{2}{c|}{$\sqrt{s}= 3000$ GeV} \\
		-- & \multicolumn{2}{c|}{($\mathcal{L}=2000$ fb$^{-1}$)} &  \multicolumn{2}{c|}{($\mathcal{L}=4000$ fb$^{-1}$)} & \multicolumn{2}{c|}{($\mathcal{L}=2000$ fb$^{-1}$)} \\
		\hline
		Cuts & S (fb) & BG (fb) & S (fb) & BG (fb) & S (fb) & BG (fb) \\
		\hline \hline
		Parton-level cuts & 4.3 & 3184 & 7.4  & 1693  & 1.3 & 449 \\
		\hline
		Pre-selection cuts & 1.4 & 1650.0 & 2.2 & 877.8 & 0.01 & 166.3 \\
		\hline
		$ M_{\ell j j} \in [85, 102] \GeV$ & 0.7 & 13.4 & -- & -- & -- & --  \\
		$ \Delta R_{jj} \in [1.8, 4]$ & 0.7 & 6.4 & -- & -- & -- & -- \\
		$ | \eta_{\ell} | \in [0.5, 1.4]$ & 0.4 & 1.9 & -- & -- & -- & -- \\
		$ M_{j j} \in [68, 95] \GeV$ & 0.3 & 0.7 & -- & -- & -- & --  \\
		\hline
		$ M_{\ell j j} \in [40, 105] \GeV$ & -- & -- & 1.8 & 4.1 & -- & --  \\
		$ M_{j j} \in [67, 90] \GeV$ & -- & -- & 1.2 & 1.5 & -- & --   \\
		\hline
		$ M_{\ell j j} \in [60, 110] \GeV$ & -- & -- & -- & -- & 0.008 & 0.06 \\
		$ M_{\ell \nu} \in [0, 500] \GeV$ & -- & -- & -- & -- & 0.008 & 0.02  \\
		$ p_{T, \ell} \in [0, 50] \GeV$ & -- & -- & -- & -- & 0.008 & 0.006  \\
		$ \Delta R_{\ell j_1} \in [0, 0.9]$ & -- & -- & -- & -- & 0.008 &  $0^\dagger$ \\
		\hline
		$S/B$ & \multicolumn{2}{c|}{0.5} & \multicolumn{2}{c|}{0.8} & \multicolumn{2}{c|}{2.8} \\
		$S/\sqrt{B}$ & \multicolumn{2}{c|}{18.9} & \multicolumn{2}{c|}{61.8}  & \multicolumn{2}{c|}{6.5} \\
		$S/\sqrt{S+B}$ & \multicolumn{2}{c|}{15.3} & \multicolumn{2}{c|}{46.1} & \multicolumn{2}{c|}{3.3} \\
		\hline
	\end{tabular}  \\
	\vspace{0.5cm}
	\caption{Cut flow Table for generalized CW model \textbf{G100}.}
	\label{tab:cutflow_lepton_colliders_generalized_model_1}
\end{table}
\begin{table}[ht]
	\vspace{1.0cm}
	\begin{tabular}{|c|c|c|c|c|c|c|}
		\hline 
		-- & \multicolumn{4}{c|}{ \textbf{G300} } & \multicolumn{2}{c|}{ \textbf{G750} } \\
		\hline
		-- & \multicolumn{2}{c|}{ $\sqrt{s}= 500$ GeV} &  \multicolumn{2}{c|}{$\sqrt{s}= 3000$ GeV} & \multicolumn{2}{c|}{$\sqrt{s}= 3000$ GeV} \\
		-- & \multicolumn{2}{c|}{($\mathcal{L}=4000$ fb$^{-1}$)} &  \multicolumn{2}{c|}{($\mathcal{L}=2000$ fb$^{-1}$)} & \multicolumn{2}{c|}{($\mathcal{L}=2000$ fb$^{-1}$)} \\
		\hline
		Cuts & S (fb) & BG (fb) & S (fb) & BG (fb) & S (fb) & BG (fb) \\
		\hline \hline
		Parton-level cuts & 5.8  & 1693 & 6.1  & 449 & 7.9 & 449 \\
		\hline
		Pre-selection cuts & 3.4 & 780.9 & 1.2 & 155.8 & 1.9 & 155.8 \\
		\hline
		$ M_{\ell j j} \in [290, 335] \GeV$ & 2.5 & 79.9 & -- & -- & -- & --  \\
		$ p_{T, \ell} \in [100, 250] \GeV$ & 2.0 & 18.0 & -- & -- & -- & -- \\
		$ p_{T, j_1} \in [70, 150] \GeV$ & 1.7 & 9.1 & -- & -- & -- & -- \\
		$ \Delta R_{\ell j j} \in [0.8, 1.5]$ & 1.5 & 4.8 & -- & -- & -- & --  \\
		$ \cancel{E_T} \in [0, 85] \GeV$ & 1.3 & 3.3 & -- & -- & -- & --  \\
		\hline
		$ M_{\ell j j} \in [250, 340] \GeV$ & -- & -- & 1.1 & 11.0 & -- & --  \\
		$  | \eta_{j_1} | \in [1.6, 2.5]$ & -- & -- & 1.0 & 4.2 & -- & --   \\
		$ \cancel{E_T} \in [0, 150] \GeV$ & -- & -- & 0.7 & 1.4 & -- & --   \\
		\hline
		$ M_{\ell j j} \in [720, 820] \GeV$ & -- & -- & -- & -- & 1.5 & 10.7 \\
		$ p_{T, j_1} \in [140, 450] \GeV$ & -- & -- & -- & -- & 1.3 & 4.7  \\
		$  | \eta_{j_1} | \in [0.6, 2.5]$ & -- & -- & -- & -- & 1.0 & 1.5 \\
		$ M_{\ell \nu} \in [0, 510] \GeV$ & -- & -- & -- & -- & 0.6 & 0.5  \\
		\hline
		$S/B$ & \multicolumn{2}{c|}{ 0.4} & \multicolumn{2}{c|}{0.5} & \multicolumn{2}{c|}{1.4} \\
		$S/\sqrt{B}$ & \multicolumn{2}{c|}{43.7} & \multicolumn{2}{c|}{26.3} & \multicolumn{2}{c|}{42.5} \\
		$S/\sqrt{S+B}$ & \multicolumn{2}{c|}{37.2} & \multicolumn{2}{c|}{21.4} & \multicolumn{2}{c|}{27.5} \\
		\hline
	\end{tabular}
	\vspace{0.5cm}
	\caption{Cut flow Table for generalized CW models: \textbf{G300} and \textbf{G750}.
		${}^\dagger$ The fact that we get 0 events is an artifact of low statistics of our background sample. When we estimate the signal significance, we used Poisson statistics and used 3 Madgraph events, which corresponds to 5.4 actual events with the integrated luminosity shown.}
	\label{tab:cutflow_lepton_colliders_generalized_model_2}
\end{table} 
%
%
%


%
%
\begin{table}[ht]
	\centering
	\begin{tabular}{|c|c|c|c|c|c|c|}
		\hline 
		\multicolumn{7}{|c|}{\textbf{U100}} \\
		\hline
		-- & \multicolumn{2}{c|}{$\sqrt{s}= 250$ GeV} &  \multicolumn{2}{c|}{$\sqrt{s}= 500$ GeV} & \multicolumn{2}{c|}{$\sqrt{s}= 3000$ GeV} \\
		-- & \multicolumn{2}{c|}{($\mathcal{L}=2000$ fb$^{-1}$)} &  \multicolumn{2}{c|}{($\mathcal{L}=4000$ fb$^{-1}$)} & \multicolumn{2}{c|}{($\mathcal{L}=2000$ fb$^{-1}$)} \\
		\hline
		Cuts & S (fb) & BG (fb) & S (fb) & BG (fb) & S (fb) & BG (fb) \\
		\hline \hline
		Parton-level cuts & 7.4 & 3184 & 12.8  & 1693  & 3.9 & 449 \\
		\hline
		Pre-selection cuts & 3.2 & 1650.0 & 5.3 & 877.8 & 0.09 & 166.3 \\
		\hline
		$ M_{\ell j j} \in [70, 140] \GeV$ & 2.3 & 163.3 & -- & -- & -- & --  \\
		$ \Delta R_{jj} \in [1.8, 3.5]$ & 1.9 & 91.5 & -- & -- & -- & -- \\
		$ M_{\ell \nu} \cancel{\in} [60, 90] \GeV$ & 1.3 & 43.0 & -- & -- & -- & --  \\
		\hline
		$ M_{\ell j j} \in [50, 150] \GeV$ & -- & -- & 4.3 & 30.8 & -- & --  \\
		$ | \eta_{j_1} | \in [0.7, 2.5]$ & -- & -- & 3.4 & 17.9 & -- & --   \\
		$ \eta_{\ell} \in [-1.4, 1.4]$ & -- & -- & 2.3 & 6.9 & -- & --   \\
		\hline
		$ M_{\ell j j} \in [80, 160] \GeV$ & -- & -- & -- & -- & 0.08 & 1.0 \\
		$ p_{T, j_1} \in [100, 400] \GeV$ & -- & -- & -- & -- & 0.06 & 0.5  \\
		$ p_{T, \ell} \in [0, 100] \GeV$ & -- & -- & -- & -- & 0.04 & 0.2  \\
		$ \eta_{\ell} \in [-2,2]$ & -- & -- & -- & -- & 0.02 & 0.04 \\
		\hline
		$S/B$ & \multicolumn{2}{c|}{0.03}  & \multicolumn{2}{c|}{0.3}  & \multicolumn{2}{c|}{0.6} \\
		$S/\sqrt{B}$ & \multicolumn{2}{c|}{9.0} & \multicolumn{2}{c|}{54.5} & \multicolumn{2}{c|}{5.2} \\
		$S/\sqrt{S+B}$ & \multicolumn{2}{c|}{8.9} & \multicolumn{2}{c|}{47.3}  & \multicolumn{2}{c|}{4.1} \\
		\hline
	\end{tabular}  \\
	\vspace{0.5cm}
	\caption{Cut flow Table for uniform CW model for \textbf{U100}.}
	\label{tab:cutflow_lepton_colliders_uniform_model_1}
\end{table}
\begin{table}[ht]
	\vspace{1.0cm}
	\begin{tabular}{|c|c|c|c|c|c|c|c|c|}
		\hline 
		-- & \multicolumn{4}{c|}{ \textbf{U400} } & \multicolumn{2}{c|}{ \textbf{U750} }  & \multicolumn{2}{c|}{ \textbf{U1000} } \\
		\hline
		-- & \multicolumn{2}{c|}{$\sqrt{s}= 500$ GeV} &  \multicolumn{2}{c|}{$\sqrt{s}= 3000$ GeV} & \multicolumn{2}{c|}{$\sqrt{s}= 3000$ GeV} & \multicolumn{2}{c|}{$\sqrt{s}= 3000$ GeV} \\
		-- & \multicolumn{2}{c|}{($\mathcal{L}=4000$ fb$^{-1}$)} &  \multicolumn{2}{c|}{($\mathcal{L}=2000$ fb$^{-1}$)} & \multicolumn{2}{c|}{($\mathcal{L}=2000$ fb$^{-1}$)} & \multicolumn{2}{c|}{($\mathcal{L}=2000$ fb$^{-1}$)} \\
		\hline
		Cuts & S (fb) & BG (fb) & S (fb) & BG (fb) & S (fb) & BG (fb) & S (fb) & BG (fb) \\
		\hline \hline
		Parton-level cuts & 0.8  & 1693 & 7.6  & 449 & 5.9 & 449 & 2.3 & 449 \\
		\hline
		Pre-selection cuts & 0.5 & 780.9 & 3.0 & 155.8 & 1.2 & 155.8 & 0.4 & 155.8 \\
		\hline
		$ M_{\ell j j} \in [380, 460] \GeV$ & 0.4 & 285.1 & -- & -- & -- & -- & -- & --  \\
		$ p_{T, j_1} \in [90, 250] \GeV$ & 0.3 & 116.6 & -- & -- & -- & -- & -- & --  \\
		$ p_{T, \ell} \in [190, 240] \GeV$ & 0.1 & 10.1 & -- & -- & -- & -- & -- & --  \\
		\hline
		$ M_{\ell j j} \in [400, 600] \GeV$ & -- & -- & 2.7 & 29.1 & -- & -- & -- & --  \\
		$  | \eta_{j_1} | \in [1.4, 2.5]$ & -- & -- & 1.9 & 5.1 & -- & --  & -- & --  \\
		$ \cancel{E_T} \in [0, 150] \GeV$ & -- & -- & 1.6 & 2.7 & -- & -- & -- & --   \\
		\hline
		$ M_{\ell j j} \in [740, 1020] \GeV$ & -- & -- & -- & -- & 1.0 & 26.6 & -- & --  \\
		$ p_{T, j_1} \in [150, 600] \GeV$ & -- & -- & -- & -- & 0.8 & 10.9 & -- & --   \\
		$ M_{\ell \nu} \in [0, 800] \GeV$ & -- & -- & -- & -- & 0.7 & 5.4 & -- & --  \\
		$ \cancel{E_T} \in [0, 130] \GeV$ & -- & -- & -- & -- & 0.6 & 3.4 & -- & --  \\
		$ \eta_{\ell} \in [-2.2, 2.2]$ & -- & -- & -- & -- & 0.4 & 1.5 & -- & --  \\
		$  | \eta_{j_1} | \in [0.4, 2.5]$ & -- & -- & -- & -- & 0.4 & 0.8 & -- & --  \\
		\hline
		$ M_{\ell j j} \in [1050, 1450] \GeV$ & -- & -- & -- & -- & -- & --  & 0.3 & 24.2 \\
		$ p_{T, j_1} \in [150, 700] \GeV$ & -- & -- & -- & --  & -- & -- & 0.3 & 11.3  \\
		$ M_{\ell \nu} \in [0, 650] \GeV$ & -- & -- & -- & --  & -- & --  & 0.2 & 3.8 \\
		$ \cancel{E_T} \in [0, 100] \GeV$ & -- & -- & -- & --  & -- & --  & 0.2 & 2.2 \\
		$ \eta_{\ell} \in [-2.2, 2.2]$ & -- & -- & -- & --  & -- & -- & 0.09 & 0.7 \\
		\hline
		$S/B$ & \multicolumn{2}{c|}{0.01} & \multicolumn{2}{c|}{0.6} & \multicolumn{2}{c|}{0.4} & \multicolumn{2}{c|}{0.1}  \\
		$S/\sqrt{B}$ & \multicolumn{2}{c|}{2.9} & \multicolumn{2}{c|}{43.5} & \multicolumn{2}{c|}{18.0}  & \multicolumn{2}{c|}{5.0}  \\
		$S/\sqrt{S+B}$ & \multicolumn{2}{c|}{2.8} & \multicolumn{2}{c|}{34.4} & \multicolumn{2}{c|}{14.9}  & \multicolumn{2}{c|}{4.7}  \\
		\hline
	\end{tabular}
	\vspace{0.5cm}
	\caption{Cut flow Table for uniform CW models: \textbf{U400}, \textbf{U750}, and \textbf{U1000}}
	\label{tab:cutflow_lepton_colliders_uniform_model_2}
\end{table} 

\clearpage

\bibliography{lit}
\bibliographystyle{jhep}

\end{document}